\documentclass[a4paper,11pt]{article}
\pdfoutput=1 

\usepackage{jcappub} 

\usepackage{verbatim}
\usepackage{epsfig}
\usepackage{graphicx}
\usepackage{booktabs}
\usepackage{multirow}
\usepackage{dcolumn}
\usepackage{amsmath}
\usepackage{mathtools}
\usepackage{amsfonts}
\usepackage{amssymb}
\usepackage{epstopdf}
\usepackage{bm}
\usepackage{braket}
\usepackage{enumitem}
\usepackage{soul}
\usepackage[table]{xcolor}

\definecolor{navyblue}{rgb}{0.0, 0.0, 0.5}
\definecolor{royalblue}{rgb}{0.25, 0.41, 0.88}
\definecolor{cadmiumgreen}{rgb}{0.0, 0.42, 0.24}
\definecolor{blue-violet}{rgb}{0.54, 0.17, 0.89}
\definecolor{darkviolet}{rgb}{0.58, 0.0, 0.83}
\definecolor{orange(colorwheel)}{rgb}{1.0, 0.5, 0.0}

\usepackage{hyperref}
\hypersetup{colorlinks=true,linkcolor=royalblue,citecolor=magenta,pdfencoding=auto}
\usepackage{pifont}
\usepackage{dcolumn}
\usepackage{colortbl}

\newcommand\ee{\end{equation}}
\newcommand\be{\begin{equation}}
\newcommand\eea{\end{eqnarray}}
\newcommand\bea{\begin{eqnarray}}

\renewcommand\[{\left[}

\newcommand{\ns}{n_{\rm s}}

\newcommand\limit[1]{#1\%\,\mathrm{CL}}

\newcommand{\dif}{\mathrm{d}}
\newcommand\ie{{\it i.e.}~}
\newcommand\eg{{\it e.g.}~}

\newcommand\eq[1]{Eq.~\eqref{eq:#1}}
\newcommand\eqsI[1]{Eqs.~\eqref{eq:#1}}
\newcommand{\eqsII}[2]{Eqs.~\eqref{eq:#1}, \eqref{eq:#2}}

\newcommand\sect[1]{Sec.~\ref{sec:#1}}
\newcommand\fig[1]{Fig.~\ref{fig:#1}}
\newcommand\tab[1]{Tab.~\ref{tab:#1}}
\newcommand{\Tr}{\mathrm{Tr}}
\renewcommand{\vec}{\bm}

\newcommand\eps{\varepsilon}

\newcommand{\D}[2]{\frac{\dif #1}{\dif #2}}
\newcommand{\prt}[2]{\frac{\partial #1}{\partial #2}}

\newcommand{\intk}[1]{\int\frac{\dif #1}{(2\pi)^3}}
\newcommand{\COS}{\mu}
\newcommand{\cH}{\mathcal{H}}
\newcommand{\vphi}{\varphi}
\newcommand{\zetal}{\zeta_\ell}
\newcommand{\zetas}{\zeta_s}
\newcommand{\kl}{k_\ell}
\newcommand{\ks}{k_s}
\newcommand{\vkl}{\vec{k}_\ell}
\newcommand{\vks}{\vec{k}_s}
\newcommand{\ds}{\frac{\dif}{\dif\log\ks}}
\newcommand{\dsh}[1]{\frac{\dif^{#1}}{\dif\log\ks^{#1}}}
\newcommand{\x}{\vec{x}}
\newcommand{\R}{\vec{r}}

\newcommand{\bx}{\bar{x}}
\newcommand{\vbx}{\bar{\vec{x}}}
\newcommand{\cs}{c_\mathrm{s}}
\newcommand{\fx}{x_F}
\newcommand{\fbx}{\vec{x}^F}
\newcommand{\fnl}{f_\mathrm{NL}}

\newcommand\vertsp{\rule[-2mm]{1mm}{0mm} &}
\newcommand\horsp{\rule[-1.5mm]{0mm}{4.125mm}}
\newcommand\morehorsp{\rule[-2.25mm]{0mm}{6mm}}
\newcommand\moremorehorsp{\rule[-2.75mm]{0mm}{8mm}}

\definecolor{magenta(process)}{rgb}{1.0, 0.0, 0.56}

\definecolor{darkspringgreen}{rgb}{0.09, 0.45, 0.27}

\definecolor{royalblue(web)}{rgb}{0.25, 0.41, 0.88}

\DeclarePairedDelimiter{\abs}{\lvert}{\rvert}

\title{How Gaussian can our Universe be?}

\author[a]{G. Cabass,}
\author[b]{E. Pajer,}
\author[c]{F. Schmidt}

\affiliation[a]{Physics Department and INFN, Universit\`a di Roma 
	``La Sapienza'', 
	P.le\ Aldo Moro 2, 00185, Rome, Italy}
\affiliation[b]{Institute for Theoretical Physics and Center for Extreme Matter and Emergent Phenomena,
	Utrecht University, 
	Princetonplein 5, 3584 CC Utrecht, The Netherlands}
\affiliation[c]{Max-Planck-Institut f\"{u}r Astrophysik, 
Karl-Schwarzschild-Str. 1, 85741 Garching, Germany}

\emailAdd{giovanni.cabass@roma1.infn.it}
\emailAdd{e.pajer@uu.nl}
\emailAdd{fabians@mpa-garching.mpg.de}

\abstract{\noindent Gravity is a non-linear theory, and hence, barring cancellations, the initial super-horizon perturbations produced by inflation 
must contain some minimum amount of mode coupling, or primordial non-Gaussianity. 
In single-field slow-roll models, where this lower bound is saturated, non-Gaussianity is controlled by two observables: the tensor-to-scalar ratio, which is uncertain by more than fifty 
orders of magnitude; and the scalar spectral index, or tilt, which is relatively well measured. It is well known that to leading and next-to-leading order in derivatives, the contributions 
proportional to the tilt disappear from any local observable, and suspicion has been raised that this might happen to all orders, allowing for an arbitrarily low amount of primordial 
non-Gaussianity. Employing Conformal Fermi Coordinates, we show explicitly that this is not the case. Instead, a contribution of order the tilt appears in local 
observables. In summary, the floor of physical primordial non-Gaussianity in our Universe has a squeezed-limit scaling of $k_\ell^2/k_s^2$, similar to equilateral and orthogonal shapes, and 
a dimensionless amplitude of order $0.1\times(n_\mathrm{s}-1)$.}

\begin{document}
\maketitle
\flushbottom

\section{Introduction}
\label{sec:intro}

\noindent As cosmological observations show no evidence of departures from Gaussian primordial perturbations, it is natural to ask: 
How Gaussian can our Universe be? If we assume primordial perturbations to be generated during inflation, we know that multi-field and higher derivative interactions typically enhance 
primordial non-Gaussianity. Setting aside these more general scenarios, we focus on the simplest model, which leads to the least amount of non-Gaussianity: canonical single-field 
slow-roll inflation. We know that inflaton self-interactions are subleading in the slow-roll expansion (see \cite{Falk:1992zg,Maldacena:2002vr,Burrage:2011hd,Pajer:2016ieg} 
for explicit calculations), so we are led to ask how small gravitational non-linearities can be. Maldacena answered this question in \cite{Maldacena:2002vr} 
computing the primordial bispectrum in comoving coordinates
\begin{equation}
\label{eq:shape_function_definition}
\begin{split}
&B_\zeta(k_1,k_2,k_{3})
\propto\frac{(\Delta^2_\zeta)^2}{(k_1k_2k_{3})^2} \bigg[(1-\ns)\,\mathcal{S}_\mathrm{loc.}(k_1,k_2,k_3) + \frac{5}{3}\eps\,\mathcal{S}_\mathrm{equil.}(k_1,k_2,k_3)\bigg]\,\,,
\end{split}
\end{equation}
where $P_\zeta(k) = k^3 \Delta^2_\zeta/2\pi^2$ is the power spectrum of curvature perturbations, $\mathcal{S}_\mathrm{loc.}$ and $\mathcal{S}_\mathrm{equil.}$ are the 
shape functions of local and equilateral non-Gaussianity, and $\ns-1$ is the scalar spectral tilt, which is given in terms of the Hubble slow-roll parameters by
\begin{equation}
\label{eq:tilt_and_HSR_parameters}
\text{$
\ns-1 = -\eta - 2\eps
$, with $\eps\equiv-\frac{\dot{H}}{H^2}$, $\eta\equiv\frac{\dot{\eps}}{H\eps}$}\,\,.
\end{equation}
The minimum size of non-Gaussianity is therefore determined by $\eps$ and the spectral tilt $\ns-1$. Although these two contributions are ``of order slow-roll'', 
there is a dramatic difference between the two. The spectral tilt is relatively well known, $\ns-1 = -0.0355 \pm 0.005$ ($\limit{95}$) \cite{Ade:2015lrj}. On the other hand, $\eps$ is uncertain 
by more than $50$ orders of magnitude: an upper bound comes from the tensor-to-scalar ratio bound $16\eps = r <0.07$ ($\limit{95}$) \cite{Array:2015xqh}, 
while a lower bound comes from conservatively assuming a reheating scale larger than a $\mathrm{TeV}$, leading to $5\times 10^{-3}\gtrsim\eps\gtrsim10^{-54}$. 
So the answer to the title of this paper can be hugely different, depending on whether it is $\eps$ or $\ns-1$ that control the minimum amount of primordial non-Gaussianity.

It was shown in \cite{Pajer:2013ana} that, to leading order in derivatives, the contribution from the local shape, of size $\ns-1$, cancels exactly for any local measurement. 
In particular, it does not contribute to the scale-dependent bias \cite{Dai:2015rda,Dai:2015jaa,dePutter:2015vga}, to the CMB bispectrum in the 
squeezed limit \cite{Creminelli:2004pv,Creminelli:2011sq}, and to the cross-correlation between CMB temperature anisotropies and spectral 
distortions \cite{Pajer:2012vz}. It is therefore natural to ask whether $\ns-1$ survives at some subleading order in derivatives or if it cancels to all 
orders, allowing primordial non-Gaussianity to be, for all practical purposes, arbitrarily small. The goal of this paper is to answer this question. 
Using Conformal Fermi Coordinates (CFC) \cite{Pajer:2013ana,Dai:2015rda,Dai:2015jaa}, we will show that a term which involves two spatial derivatives of $\zeta$ and is proportional to $\ns-1$ survives 
in local observables and therefore appears in the appropriately defined curvature bispectrum. 

To put our result into context, we stress two main points. First, the original motivation for our investigation was the widespread suspicion, put forward in \cite{Alvarez:2014vva}, 
that some general argument might exist to guarantee the complete cancellation of any term 
proportional to $\eta$ (and therefore to the tilt). After all, it is $\eps$ that controls the departure from an exact de Sitter spacetime (see \eq{riemann_dS}), in which case, 
following the argument sketched in \cite{Alvarez:2014vva}, non-Gaussianity should vanish. Our explicit calculation shows that this suspicion is unfounded. 
We also clarify how the survival of $\ns-1$ is indeed expected when considering the de Sitter limit. Second, even though we compute the bispectrum 
of primordial curvature perturbations on a constant-proper-time hypersurface at the end of inflation, as opposed to some late-time observable such as the CMB or 
galaxy bispectrum, our result has a direct and transparent physical implication. 

Recall that curvature perturbations, and hence their correlators, are conserved 
until they re-enter the (largest) sound horizon of any relevant component (matter, radiation, etc.). As an example, consider then matter domination, when the 
sound horizon is parametrically smaller than the Hubble radius. Two short and one long mode that enter the Hubble horizon during this epoch 
still possess the primordial correlation we compute here as long as they are larger than the sound horizon, and 
this coupling is in principle observable, as we will discuss in \sect{observations}. In practice of course we are 
interested in modes that enter also during radiation domination and we observe non-conserved density perturbations as opposed to conserved 
curvature perturbations. Many evolution and projection effects then need to be added to our result. Nevertheless, the example above highlights that our 
result describes a physical and in principle measurable late-time correlation. Connection to observations will be further discussed in \sect{observations}. 

The rest of the paper is organized as follows. In \sect{cfc} we construct the CFC frame for single-field slow-roll inflation; in \sect{bispectrum_transformation} 
and \sect{field_theory} we compute the local bispectrum in CFC; in \sect{argument} we discuss why we expect $\eta$ to also be locally observable, in contrast to 
what was argued in \cite{Alvarez:2014vva}, and briefly describe the case where the inflaton speed of sound $c_\mathrm{s}$ is different from $1$. Finally, we derive 
our conclusions in \sect{discussion}. We collect the technical details in \sect{details_for_cfc} (about the CFC construction), \sect{appendix_active} (about the transformation 
of the curvature perturbation $\zeta$ from comoving coordinates to CFC), and \sect{bispectrum_fourier} (about the bispectrum in Fourier space). 
In \sect{small_sound_speed} we briefly describe the simplifications in the calculation of the CFC bispectrum when $\cs\ll 1$. 

\vspace*{0.5\baselineskip} 

\noindent\textbf{Notation and conventions} We use natural units $c = \hbar = 1$, and the ``mostly plus'' metric signature. As we did in the introduction above, we 
use $\zeta$ (not $\mathcal{R}$) to define the comoving curvature perturbation, following \cite{Maldacena:2002vr}. In the remainder of the paper, we work in units 
where the reduced Planck mass $M^2_\mathrm{P}\equiv 1/8\pi G_\mathrm{N} = 1$, unless it is explicitly said otherwise. It can be reintroduced easily with dimensional 
analysis in the final results, if needed.

\section{CFC coordinates in canonical single-field inflation}
\label{sec:cfc}

\begin{figure*}[!hbt]
\begin{center}
\includegraphics[width=0.85\columnwidth]{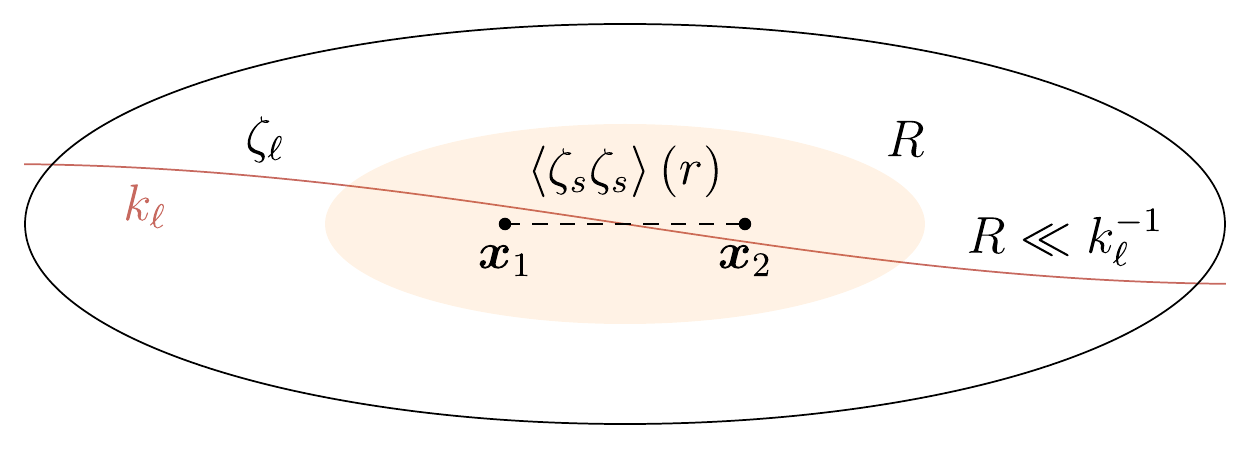} 
\end{center}
\caption{\footnotesize{Squeezed limit of $\zeta = \zetal + \zetas$ in real space: we compute how the correlation function of $\zetas$ (which we call $\braket{\zetas\zetas}(r)$, 
where $r\equiv\abs{\vec{x}_1-\vec{x}_2}$) depends on the long-wavelength fluctuation $\zetal$. We can expand $\zetal$ in a Taylor series, since it is slowly varying inside $R$: any 
point inside of $R$ is as good as the other for the expansion \cite{Creminelli:2011rh,Pajer:2013ana}, so we will choose the middle point $\vec{x}_c\equiv(\vec{x}_1+\vec{x}_2)/2$ for 
simplicity. We also stress that the choice of $R$ is immaterial in the squeezed limit, the only real requirement being that $\kl\ll\ks$ \cite{Wagner:2015gva}.}} 
\label{fig:figure_region}
\end{figure*}

\noindent When we referred to ``local measurements'' in the introduction above, we meant in particular the response of short-wavelength 
perturbations ($k_1\sim k_2\sim\ks$) to the presence of long-wavelength ones $k_3\sim\kl\ll\ks$ (squeezed limit). As shown, \emph{e.g.}, in \cite{Senatore:2012wy},\footnote{See 
its Sec.~2.} 
the squeezed limit of correlation functions of $\zeta$ in Fourier space corresponds to looking at how perturbations $\zetas$ which are defined in a region 
of size $R\gtrsim\ks^{-1}$ are correlated with perturbations $\zetal$ of wavelength $\kl^{-1}\gg R$, \ie that are 
almost constant in the region $R$ (see \fig{figure_region}). This correlation between long and short modes is expected, since the long modes will affect 
the dynamics of the short modes, modifying the background over which they evolve: up to second order in gradients, the long-wavelength perturbation can be 
reabsorbed in the FLRW background, while at $\mathcal{O}(\kl^2)$ it adds curvature to the ``separate universe'' of size $\sim\kl^{-1}$ and modifies its expansion 
history \cite{Baldauf:2011bh,Creminelli:2013cga}. Maldacena's consistency relation is just a statement of the fact that these effects are suppressed by how 
much the long mode is outside the horizon at a given time (for primordial correlations this saturates at $\kl^2/\ks^2$, $\ks\sim aH\equiv\cH$ being the moment 
when short modes freeze out).

We can see this in the following way. We start by asking ourselves what a local observer with proper $4$-velocity $U^\mu$ in the separate universe of \fig{figure_region}, freely falling 
in the background perturbed by the long-wavelength mode $\zetal$, can measure during inflation. First, she naturally sets the time coordinate to match 
what is measured by her clock (\ie by her proper time $\equiv t_F$) and uses it to define surfaces of constant time. The (non-rotating) 
spatial coordinate axes of her local laboratory frame emanate from her worldline along geodesics. The resulting coordinate system ($x_F^\mu$) depends on the 
worldline of $U^\mu$: timelike and spacelike coordinates are defined in such a way that the distance of a point from the worldline 
is given by $\eta_{\mu\nu}\Delta x^\mu_F \Delta x^\nu_F$, with higher order corrections in $\Delta\vec{x}_F$ that encode how spacetime deviates from flatness. 
This local coordinate system is known as Fermi Normal 
Coordinates (FNC) \cite{Manasse:1963zz,Baldauf:2011bh,Creminelli:2012ed,Senatore:2012wy,Senatore:2012ya}. 

However, space is also expanding, as determined locally by how a small sphere of test particles carried on the worldline changes in volume. This change is encoded by the geodesic expansion $\nabla_\mu U^\mu$. By introducing a local FLRW scale factor $a_F$, the spatial coordinates can account for the fact that $\nabla_\mu U^\mu\neq 0$: 
the distance of a point from the worldline is then given by $-\Delta t_F^2 + a^2_F\abs{\Delta\vec{x}_F}^2$, where $a_F$ is the integral over the local expansion rate $\nabla_\mu U^\mu = 3 H_F$. 
The important point is that spatial geodesics are \emph{still} used to define spatial distances, the only difference with the previous case being the fact that the overall 
expansion of space has now been factored out. Higher-order corrections in $\Delta\vec{x}_F$ to the 
distance between points would now encode the intrinsic curvature of spatial slices. 

This generalization of the FNC is called Conformal Fermi 
Coordinates (CFC) \cite{Pajer:2013ana,Dai:2015rda,Dai:2015jaa}: they are the coordinates that a local observer uses to describe physics in an \emph{expanding} 
universe. They are naturally suited to the case where there is a separation of scales, such as the one described in \fig{figure_region}: an observer who has access only to 
scales $\sim 1/\ks$ treats the long mode as an effective background within which the short modes evolve,\footnote{It is clear that this picture, during inflation, can hold only if we stop 
at quadratic order in gradients of the long mode: at higher order we cannot neglect the quantum nature of perturbations and treat them as a classical background. To see this, it is 
enough to think about the de Sitter mode functions $\zeta(\tau,k) = \zeta(0,k)(1+ik\tau)e^{-ik\tau}$: for $k\to0$, the term of $\mathcal{O}(k^3)$ picks up a factor of $i$.} and then 
looks at what is the power spectrum of the latter in this background, which she describes through CFC. This coordinate system makes explicit that the separate universe is 
an unperturbed FLRW universe (the corrections to the expansion history coming from $H_F\neq H$ are of order of the time dependence of $\zetal$, which starts at 
order $\partial^2\zetal$ in single-field inflation): deviations from this picture enter only at second order in spatial gradients of $\zetal$. Hence, the first non-zero, physical 
coupling between short and long modes that a local observer can measure appears at quadratic order in the momentum of the long mode. At this order, if the CFC power 
spectrum of $\zetas$ in presence of $\zetal$ does not vanish for $\eps\to 0$ on super-Hubble scales (we will show later that 
the difference between constant $t$ surfaces and constant $t_F$ surfaces goes to zero as the Hubble radius decreases), we conclude that the ``gravitational floor'' of 
non-Gaussianities from inflation is of order of the tilt $\ns-1$.

\subsection{Construction of Conformal Fermi Coordinates}
\label{sec:cfc_construction}

\noindent As we explained above, CFC coordinates $x_F = (\tau_F, \vec{x}_F)$ for a geodesic observer $U^\mu \equiv (e_0)^\mu$ are constructed in a similar way to 
Fermi Normal Coordinates, the difference being that around the observer's geodesic the metric looks approximately as FLRW (not Minkowski). The deviations 
from FLRW are of order $\abs{\vec{x}_F}^2\kl^2\zetal$, 
instead of $\abs{\vec{x}_F}^2H^2$ as in the FNC case. 
The construction goes as follows:
\begin{enumerate}[leftmargin=*]
\item we construct an orthonormal tetrad $(e_\nu)^\mu$, parallel transported along the central geodesic $P(t_F)$ of the observer $(e_0)^\mu$ ($t_F$ being the observer's proper time);
\item given a spacetime scalar $a_F(x)$, we define a conformal proper time $\tau_F$ by 
\begin{equation}
\label{eq:conformal_proper_time}
\dif\tau_F = a^{-1}_F(P(t_F))\dif t_F\,\,,
\end{equation}
and we choose $\tau_F$ as the time coordinate (often replacing $P(t_F(\tau_F))$ with just $P$ to simplify the notation). 
This allows us to define surfaces of constant $\tau_F$, spanned by space-like conformal geodesics (\ie geodesics of the conformal 
metric $\tilde{g}_{\mu\nu}(x)\equiv a^{-2}_F(x) g_{\mu\nu}(x)$) originating from the central geodesic;
\item this construction of surfaces of constant $\tau_F$ also gives us spatial coordinates $x^i_F$. More precisely:
\begin{itemize}
\item one defines the central geodesic to have coordinates $x_F = (\tau_F,\vec{0})$;
\item one takes the family $\gamma(\tau_F;\alpha^i,\lambda)$ of 
geodesics of the conformal metric with affine parameter 
$\lambda = 0$ at $P$, and tangent vector given by $\alpha^i (e_i)^\mu_P$;
\item the point $Q$ with coordinates $(\tau_F,\vec{x}_F)$ is then identified with $\gamma(\tau_F;\beta^i,\lambda_Q)$, where
\begin{subequations}
\label{eq:cfc_general_params}
\begin{align}
&\lambda_Q = \delta_{ij}x^i_Fx^j_F\,\,, \label{eq:cfc_general_params-1} \\
&\beta^i = \frac{a_F(P) x^i_F}{\sqrt{\delta_{ij}x^i_Fx^j_F}}\,\,; \label{eq:cfc_general_params-2}
\end{align}
\end{subequations}
\item with the exponential map we can then construct the coordinate transformation from global coordinates ($x$) to CFC coordinates ($x_F$) 
as a power series in $x_F^i$. Rescaling $\lambda$ so that it runs from $0$ to $1$, \ie 
$\beta^i = a_F(P)x^i_F$, this power series reads as
\begin{equation}
\label{eq:cfc_exponential_map}
\text{$x^\mu(x_F) = c^\mu_0(\tau_F) + \sum_{n=1}^{+\infty} c^\mu_n(\tau_F,\vec{x}_F)$, with $c^\mu_{n} (x_F) = \mathcal{O}[(x^i_F)^n]$ for $n\geq1$}\,\,.
\end{equation}
We see that $c^\mu_0(\tau_F)$ is simply given by 
$x^\mu(P)$, the coordinates of the central geodesic evaluated at $t_F(\tau_F)$, and can be computed once one 
knows $a_F$ and $(e_0)^\mu$. The tangent vector on $P$, \ie $c^\mu_1(\tau_F,\vec{x}_F)$, is then given by
\begin{equation}
\label{eq:cfc_tangent_vector}
c^\mu_1(\tau_F,\vec{x}_F) = a_F(P) (e_i)^\mu_P x^i_F\,\,.
\end{equation}
Higher order coefficients 
are computed recursively by solving the geodesic equation for the conformal metric: we refer to \cite{Dai:2015rda,Dai:2015jaa} and to \sect{details_for_cfc} for details.
\end{itemize}
\end{enumerate}

The resulting metric has the form
\begin{equation}
\label{eq:cfc_metric_general}
\text{$g^F_{\mu\nu}(x_F) = a_F^2(\tau_F)[\eta_{\mu\nu} + h^F_{\mu\nu}(x_F)]$, with $h^F_{\mu\nu}(x_F) = \mathcal{O}[(x^i_F)^2]$}\,\,.
\end{equation}
More precisely, stopping at order $(x^i_F)^2$, we have \cite{Dai:2015rda,Dai:2015jaa}
\begin{subequations}
\label{eq:cfc_metric_general-components_riemann}
\begin{align}
&h^F_{00}(x_F) = -\tilde{R}^F_{0k0l}|_Px^k_Fx^l_F\,\,, \label{eq:cfc_metric_general-components_riemann-1} \\
&h^F_{0i}(x_F) = -\frac{2}{3}\tilde{R}^F_{0kil}|_Px^k_Fx^l_F\,\,, \label{eq:cfc_metric_general-components_riemann-2} \\
&h^F_{ij}(x_F) = -\frac{1}{3}\tilde{R}^F_{ikjl}|_Px^k_Fx^l_F\,\,, \label{eq:cfc_metric_general-components_riemann-3}
\end{align}
\end{subequations}
where $\tilde{R}^F_{\mu\rho\nu\sigma}$ is the Riemann tensor of the conformal metric in CFC coordinates, and indices have been lowered 
with the conformal metric. In terms of global coordinates, $\tilde{R}^F_{\mu\rho\nu\sigma}|_P$ is
\begin{equation}
\label{eq:riemann_global_cfc}
\tilde{R}^F_{\mu\rho\nu\sigma}|_P = \tilde{R}_{\alpha\beta\gamma\delta}|_P(\tilde{e}_\mu)_P^\alpha(\tilde{e}_\rho)_P^\beta(\tilde{e}_\nu)_P^\gamma(\tilde{e}_\sigma)^\delta_P\,\,,
\end{equation}
where on the central geodesic the CFC coordinate vectors are given by $(\tilde{e}_\nu)^\mu_P = a_F(P)(e_\nu)^\mu_P$. 

When compared to the Fermi Normal Coordinates construction, CFC need one additional ingredient to 
determine the metric perturbations $h_{\mu\nu}^F$, \ie the scalar $a_F(x)$ computed along the central geodesic.\footnote{We also need its derivatives 
along the central geodesic, since these will enter in $\tilde{R}^F_{\mu\rho\nu\sigma}|_P$: we refer to \cite{Dai:2015rda,Dai:2015jaa} for a more detailed review.} 
The idea is to absorb the leading contributions to the spacetime curvature in this scale factor $a_F$, and make then the Riemann tensor of $\tilde{g}_{\mu\nu}$ as simple 
as possible. In \cite{Dai:2015rda} it is shown how this is achieved by defining $a_F(x)$ from the local \mbox{expansion rate}
\begin{equation}
\label{eq:a_F_local_hubble}
\D{\log a_F(P)}{t_F} = \frac{1}{a_F(\tau_F)}\D{\log a_F(P)}{\tau_F} = \frac{\nabla_\mu U^\mu|_P}{3}\,\,.
\end{equation}

\subsection{Residual coordinate freedom}
\label{sec:gauge_freedom}

\noindent This construction, even after having fixed the geodesic $U^\mu(t_F)$ and the choice of $a_F$, has two residual ``gauge'' freedoms that leave 
$h_{00}^F$ and $h_{0i}^F$ invariant at $\mathcal{O}[(x^i_F)^2]$: 
\begin{itemize}[leftmargin=*]
\item it is possible to perform a coordinate transformation
\begin{subequations}
\label{eq:spatial_gauge_transform_cfc}
\begin{align}
&\tau_F \to \tau_F\,\,, \label{eq:spatial_gauge_transform_cfc-1} \\
&x^i_F \to x^i_F(y_F) = y^i_F + \frac{A^i_{jkl}(\tau_F)}{6}y^j_Fy^k_Fy^l_F\,\,, \label{eq:spatial_gauge_transform_cfc-2}
\end{align}
\end{subequations}
where $A^i_{jkl}(\tau_F)$ is fully symmetric w.r.t. $j$, $k$, $l$. 
This transformation does not affect $a_F$, but changes $h^F_{\mu\nu}$ via
\begin{subequations}
\label{eq:spatial_gauge_transform_cfc-h_change}
\begin{align}
&h^F_{00}(x_F) \to h^F_{00}
(y_F)\,\,, \label{eq:spatial_gauge_transform_cfc-h_change-1} \\
&h^F_{0i}(x_F) \to h^F_{0i}
(y_F)\,\,, \label{eq:spatial_gauge_transform_cfc-h_change-2} \\
&h^F_{ij}(x_F) \to h^F_{ij}
(y_F) + 
A_{(ij)kl}(\tau_F)\,y^k_Fy^l_F\,\,, \label{eq:spatial_gauge_transform_cfc-h_change-3}
\end{align}
\end{subequations}
where the indices of 
$A^i_{jkl}(\tau_F)$ are lowered with the conformal metric. It is important to stress that, up to including order $(y^i_F)^2$, coordinate 
lines $\vec{y}_F = \lambda\,\vec{\beta}_F$ are \emph{still} geodesics of the conformal metric; 
\item one can rescale $a_F$ by a constant $a_F(\tau) \to c\,a_F(\tau)$: it 
comes from the fact that we defined it through the local Hubble rate for the observer $U^\mu$, so we still have the 
freedom of choosing the integration constant when we integrate \eq{a_F_local_hubble} along the central geodesic. 
\end{itemize}

This construction holds for any spacetime: there is no need of expanding the metric in perturbations around a given background. However, we will specialize to the case 
of a perturbed FLRW spacetime in the following sections. For this reason, we defer the discussion of these two residual transformations to the next sections, where 
working in perturbation theory will allow us to fix them in a much easier way.

\subsection{From comoving to CFC coordinates}
\label{sec:cfc_metric_inflation}

\noindent The main goal of this and the following sections is to construct explicitly the change from the global to the CFC frame, constructed 
for the long-wavelength part of the metric: this will allow us to 
find the effect that a long-wavelength perturbation $\zetal$ 
has on short modes $\zetas$. 
The construction 
will follow closely the one presented in \cite{Pajer:2013ana}, the main difference 
being the fact that 
we will go up to order $k_\ell^2/k_s^2$ in the gradient expansion. 
Here we provide the outline of the calculation, while the details are collected in \sect{details_for_cfc}.

We work in a perturbed FLRW spacetime $g_{\mu\nu} = a^2(\eta_{\mu\nu} + h_{\mu\nu})$: 
more precisely we consider the comoving gauge \cite{Maldacena:2002vr}, where the inflaton perturbations $\vphi$ are set to zero 
and the metric is given by (neglecting tensor modes) 
\begin{subequations}
\label{eq:comoving_metric}
\begin{align}
&\text{$g_{00} = a^2(-1-2N_1)$, with $N_1 = \frac{\partial_0\zeta}{\mathcal{H}}$}\,\,, \label{eq:comoving_metric-1} \\
&\text{$g_{0i} = a^2N_i = a^2\partial_i\psi$, with $\psi = -\frac{\zeta}{\mathcal{H}} + \eps\partial^{-2}\partial_0\zeta$}\,\,, \label{eq:comoving_metric-2} \\
&g_{ij} = a^2e^{2\zeta}\delta_{ij}\approx a^2(1+2\zeta)\delta_{ij}\,\,. \label{eq:comoving_metric-3}
\end{align}
\end{subequations}
Since we are interested in three-point functions, we restricted to linear order in the lapse and shift 
constraints \cite{Maldacena:2002vr,Chen:2006nt,Pajer:2016ieg}. We can now split 
$\zeta$ in a long- and short-wavelength part, $\zeta(x) = 
\zetas(x) + \zetal(x)$: because we are interested in the bispectrum only, it will be sufficient to consider the linear response of 
the short-scale modes to the coordinate transformation (that is, we can work at linear order in $\zetal$). 
Now, given that the background is FLRW, we can 
straightforwardly write down the (normalized) time-like geodesic congruence $U^\mu$ 
as \cite{Dai:2015rda,Dai:2015jaa} 
\begin{equation}
\label{eq:U_mu-flrw}
U^\mu = (e_0)^\mu = a^{-1}\bigg(1 + \frac{h_{00}}{2},V^i\bigg) = a^{-1}(1 - N_1,V^i)\,\,,
\end{equation}
where the first order perturbations $V^i$ are the peculiar velocities of the observers $U^\mu$. Neglecting vorticity (which is not 
sourced in single-field models), the corresponding spatial vectors of the tetrad are \cite{Dai:2015rda,Dai:2015jaa}
\begin{equation}
\label{eq:spatial_tetrad-flrw}
(e_i)^\mu = a^{-1}\bigg(V_i + h_{0i},\delta^j_i-\frac{h^j_i}{2}\bigg) = a^{-1}(V_i + N_i,[1-\zeta]\delta^j_i)\,\,,
\end{equation}
where we raise and lower latin indices with $\delta_i^j$. Since the tetrad $(e_\nu)^\mu$ is parallel transported along the central geodesic, one can 
show that the peculiar velocities must obey the equation
\begin{equation}
\label{eq:equation_velocities}
\partial_0V^i +\mathcal{H}V^i = -\partial^iN_1 - \partial_0N^i - \mathcal{H}N^i\,\,.
\end{equation}
Finally, one can use 
the relation $H_F = \nabla_\mu U^\mu/3$ to find the expression for the CFC scale factor $a_F$: at linear order 
in perturbations, one has that (see, \emph{e.g.}, \sect{cfc_exponential_map_at_linear_order}) 
\begin{equation}
\label{eq:a_F_over_a-A}
\frac{a_F(P)}{a(P)} = 1 + C_{a_F}(\tau_\ast,\vec{x}_c(\tau_\ast)) + \int_{\tau_\ast}^\tau\dif s\,\bigg(\partial_0\zeta(s,\vec{x}_c(s)) + \frac{1}{3}\partial_i V^i(s,\vec{x}_c(s))\bigg)\,\,,
\end{equation}
where both 
l.h.s. and 
r.h.s. of this equation are computed in global coordinates along the central geodesic ($\vec{x}_c(\tau)$). 
We have defined $\tau_\ast$ as the initial time in the integration of \eq{a_F_local_hubble}, while $C_{a_F}(\tau_\ast,\vec{x}_c(\tau_\ast))$ is an arbitrary constant
which we treat as first order in perturbations. This corresponds to the
freedom to rescale $a_F$ by a constant, as mentioned in the previous section. 
The last step is to solve 
the geodesic equation for the peculiar velocities. We can do it by defining $F^i\equiv V^i+N^i$: the solution for $F_i=\partial_i\digamma$ then reads as 
\begin{equation}
\label{eq:geodesic_eq_F_solved}
\digamma(x) = e^{-\int_{\tau_\ast}^\tau\dif s\,\mathcal{H}(s)}\bigg[\tau_\ast 
C_{\digamma}(\tau_\ast,\vec{x})-\int_{\tau_\ast}^\tau\dif s\,e^{\int_{\tau_\ast}^s\dif w\,\mathcal{H}(w)}N_1(s,\vec{x})\bigg]\,\,,
\end{equation}
where $C_{\digamma}(\tau_\ast,\vec{x})$ is a second integration constant (which we multiply by $\tau_\ast$ for convenience). 
$C_{\digamma}$ corresponds to an initial relative velocity of the geodesic (which, as we can see, decays on super-Hubble scales) considered with respect to comoving observers. 

It is now straightforward to show 
that on the central geodesic (\ie for $\vec{x}_F = \vec{0}$) we have 
\begin{equation}
\label{eq:x_mu_on_central_geodesic}
\text{$x^\mu(\tau_F,\vec{0}) = x^\mu_F + \xi^\mu(\tau_F,\vec{0})$, with $\xi^\mu(\tau_F,\vec{0}) = \mathcal{O}(\zetal)$}\,\,,
\end{equation}
so in Eqs.~\eqref{eq:a_F_over_a-A}, \eqref{eq:geodesic_eq_F_solved} we can neglect the shift in the arguments of first order 
perturbations (\ie we can take $\vec{x}_c(\tau) = \vec{0}$, $\tau = \tau_F$). This allows to write down the full exponential map 
at first order in 
the long-wavelength perturbations. If we denote by $\tilde{\Gamma}$ the Christoffel coefficients of the conformal 
metric (collected in \tab{conformal_christoffels}), and by $(a_F/a)|_\ell$ the terms of order $\zetal$ in \eq{a_F_over_a-A}, the final result at 
$\mathcal{O}[(x^i_F)^3]$ is 
equal to 
\begin{equation}
\label{eq:full_transformation-main_text}
\begin{split}
&x^\mu(\tau_F,\vec{x}_F) = x^\mu_F + \xi^\mu(\tau_F,\vec{0}) + A^\mu_i(\tau_F)\, x^i_F 
+ B^\mu_{ij}(\tau_F)\, x_F^i x_F^j + C^\mu_{kij}(\tau_F)\, x_F^i x_F^j x_F^k\,\,,
\end{split}
\end{equation}
where the coefficients of the expansion are given by 
\begin{subequations}
\label{eq:ABC_definition}
\begin{align}
&\xi^\mu(\tau_F,\vec{0}) = \begin{cases}
\int_{\tau_\ast}^{\tau_F}\dif s\big[
(a_F/a)(\tau_F,\vec{0})|_\ell - N_1(s,\vec{0})\big]
&\text{ for $\mu = 0\,\,,$} \\[1em]
\int_{\tau_\ast}^{\tau_F}\dif s\,V^l(s,\vec{0})
&\text{ for $\mu = l\,\,,$}
\end{cases}
\label{eq:ABC_definition-1} \\
&A^\mu_i(\tau_F) = \begin{cases}
F_i(\tau_F,\vec{0})&\text{ for $\mu = 0\,\,,$} \\
\big[
(a_F/a)(\tau_F,\vec{0})|_\ell - 
\zeta(\tau_F,\vec{0})\big]\delta^l_i&\text{ for $\mu = l\,\,,$}
\end{cases}
\label{eq:ABC_definition-2} \\
&B^\mu_{ij}(\tau_F) = -\frac{1}{2}\tilde{\Gamma}^\mu_{ij}(\tau_F,\vec{0})\,\,, \label{eq:ABC_definition-3} \\
&C^\mu_{kij}(\tau_F) = -\frac{1}{6}\partial_k\tilde{\Gamma}^\mu_{ij}(\tau_F,\vec{0})\,\,. \label{eq:ABC_definition-4}
\end{align}
\end{subequations}
Notice that 
$a_F$ never appears by itself. Only $\xi^0(\tau_F,\vec{0})$ and $(a_F/a)|_\ell$ appear. 

We can now 
fix the additional freedoms in the CFC construction 
(namely, the choice of $\tau_\ast$, the constants $C_{a_F}$ and $C_\digamma$, and the 
possibility of changing spatial coordinates without changing the time-time and time-space components of the metric). We start with the 
choice of initial time (noting that the initial time appears always in quantities that are already first order in the long mode): 
we are trying to absorb the effect that long-wavelength modes $\zetal$ have on short modes $\zetas$ through a 
change of coordinates. In order to do this, we must be able to treat them as classical, so we have to start defining the CFC after they have long exited the horizon. Then, we could 
choose $\tau_\ast$ such that $\cH(\tau_\ast) = \ks$, where $\ks$ is the typical wavelength of the short-scale $\zetas$. However, it is much 
simpler to choose as ``initial'' time $\tau_\ast\to 0^-$ (the end of inflation), when all modes of interest have left the horizon and $\zeta$ has 
become constant, mirroring what has been done in \cite{Dai:2015jaa}. 
This fixes the 
lower limit in the various integrals 
that define $(a_F/a)|_\ell$, the peculiar velocity potential $\digamma$, and time shift along the central geodesic $\xi^0(\tau_F,\vec{0})$. 
Now, the upper limit will also be taken to be $\tau_F\to 0^-$, since we are interested in the super-Hubble limit of correlation functions. This will simplify a lot 
the calculation, since many time integrals will not contribute. 

Then, as shown in \sect{conformal_riemann_and_cfc_long_metric}, this CFC construction 
gives 
\begin{subequations}
\label{eq:cfc_metric_general-components_riemann-explicit-main_text}
\begin{align}
&
h^F_{00} = -x_F^k x_F^l\bigg(\partial_k\partial_l - \frac{\delta_{kl}}{3}\partial^2\bigg)(N_1 + \partial_0\psi + \mathcal{H}\psi)\,\,, \label{eq:cfc_metric_general-components_riemann-explicit-main_text-1} \\
&
h^F_{0i} = \frac{2}{3} x_F^k x_F^l\big[\eps\cH^2(\delta_{kl}F_i - \delta_{ki}F_l)\big]\,\,, \label{eq:cfc_metric_general-components_riemann-explicit-main_text-2} \\
&
h^F_{ij} = -\frac{1}{3} x_F^k x_F^l\bigg[\frac{2}{3}\mathcal{H}(\partial_m V^m)T_{ijkl} + S_{ijkl}(\zeta + \mathcal{H}\psi)\bigg]
\,\,, \label{eq:cfc_metric_general-components_riemann-explicit-main_text-3}
\end{align}
\end{subequations}
where all terms on the r.h.s. are evaluated on the central geodesic (\ie at $(\tau_F,\vec{0})$), and the tensors $T_{ijkl}$, $S_{ijkl}$ are given by 
\begin{subequations}
\label{eq:tensors_for_cfc_metric-main_text}
\begin{align}
&T_{ijkl} = \delta_{il}\delta_{kj} - \delta_{ij}\delta_{kl}\,\,, \label{eq:tensors_for_cfc_metric-main_text-1} \\
&S_{ijkl} = \delta_{il}\partial_j\partial_k - \delta_{kl}\partial_i\partial_j + \delta_{kj}\partial_i\partial_l - \delta_{ij}\partial_l\partial_k\,\,. \label{eq:tensors_for_cfc_metric-main_text-2}
\end{align}
\end{subequations}
We can then use the additional 
freedom in the definition of spatial coordinates to bring the spatial part 
$h^F_{ij}$ in conformal Newtonian form, following \cite{Dai:2015jaa}. In \sect{residual_gauge_freedoms_and_gauge_fixing} we show that, at linear order in 
perturbations, the coordinate transformation of \eqsI{spatial_gauge_transform_cfc} amounts to subtracting
the tensor $A^l_{kij}(\tau_F,\vec{0})$ from $\partial_k\tilde{\Gamma}^l_{ij}$ in \eq{ABC_definition-4}. $h^F_{ij}$, correspondingly, transforms as in \eq{spatial_gauge_transform_cfc-h_change-3}. 
We perform this coordinate change with
\begin{equation}
\label{eq:A_lkij-main_text}
\begin{split}
&A^l_{kij} = -\frac{1}{6}K_F (\delta^l_k\delta_{ij} + \delta^l_i\delta_{jk}+\delta^l_j\delta_{ki}) \\
&\hphantom{A^l_{kij} =} + \frac{1}{9}(\delta^l_k\delta_{ij} + \delta^l_i\delta_{jk}+\delta^l_j\delta_{ki})\partial^2(\zeta + \cH\psi)\\
&\hphantom{A^l_{kij} =} - \frac{2}{3}(\delta^l_k\partial_i\partial_j + \delta^l_i\partial_j\partial_k + \delta^l_j\partial_k\partial_i)(\zeta + \cH\psi)\\
&\hphantom{A^l_{kij} =} + \frac{1}{3}(\delta_{ij}\partial^l\partial_k + \delta_{jk}\partial^l\partial_i + \delta_{ki}\partial^l\partial_j)(\zeta + \cH\psi)\,\,,
\end{split}
\end{equation}
where we have defined $K_F$ as
\begin{equation}
\label{eq:K_F_curvature-main_text}
K_F = -\frac{2}{3}\big[\partial^2(\zeta + \cH\psi) + \cH\partial_m V^m\big] = -\frac{2}{3}(\partial^2\zeta + \cH\partial^2\digamma)\,\,.
\end{equation}
After this final change of coordinates, the spatial metric $g_{ij}$ becomes
\begin{equation}
\label{eq:spatial_metric-stereographic-main_text}
\text{$g^F_{ij} = a^2_F
\Bigg(\frac{1 + x_F^k x_F^l\mathcal{D}_{kl}(\zeta + \mathcal{H}\psi)
}{\Big(1 + \frac{K_F\abs{\vec{x}_F}^2}{4}\Big)^2}\Bigg)\delta_{ij}$, with $\mathcal{D}_{kl}\equiv\partial_k\partial_l - \frac{\delta_{kl}}{3}\partial^2$}\,\,,
\end{equation}
which, combined with \eqsII{cfc_metric_general-components_riemann-explicit-main_text-1}{cfc_metric_general-components_riemann-explicit-main_text-2}, shows that the 
final result for the spatial metric is that of a curved FLRW metric plus tidal corrections. This form of the metric makes it clear that the scalar curvature of constant-proper-time 
slices of the observer is $\propto K_F/a^2_F$ and, as we will see in \sect{field_theory}, 
can be used to calculate the CFC bispectrum directly at the level of the action. Besides, as we will discuss in more detail in \sect{observations}, it will allow us to connect 
our result to the late-time evolution.

Finally, we can fix the constants $C_{a_F}$ and $C_\digamma$. We start from $C_{a_F}$: following \cite{Dai:2015rda,Dai:2015jaa}, we fix it by 
imposing that, at 
$\tau_F\to\tau_\ast$, the local scale factor-proper time relation is the same as that of the unperturbed background cosmology, \ie we require that 
\begin{equation}
\label{eq:fix_constant_a_F-main_text}
\lim_{\tau_F\to\tau_\ast} a_F(\tau_F) = a(\tau_\ast)\,\,. 
\end{equation}
In \sect{residual_gauge_freedoms_and_gauge_fixing} we prove that 
taking $C_{a_F} = 0$ satisfies this equality. 
We then move to $C_\digamma$, whose gradient is the initial peculiar velocity of the CFC observers. From \eq{geodesic_eq_F_solved}, we see that such initial 
velocity will decay as $1/\cH$: therefore, we can put it to zero in our treatment, since we neglect decaying modes throughout. 
In this way, we also see from \eq{cfc_metric_general-components_riemann-explicit-main_text-2} that 
the effect of a long $\zetal$ on the difference between hypersurfaces of constant $\tau$ and constant $\tau_F$ (encoded in the difference between $\tau$ and $\tau_F$ away from 
the central geodesic, which generates a non-zero $h^F_{0i}$) is of order $\kl^3$. 

With these choices for $C_{a_F}$ and $C_\digamma$, and straightforward manipulation of 
the 
lapse and shift constraints $N_1$ and $\psi$, the metric perturbations $h^F_{\mu\nu}$ 
become 
\begin{subequations}
\label{eq:cfc_metric_general-components_riemann-explicit-slow_roll}
\begin{align}
&h^F_{00}(\tau_F,\vec{x}_F) = -x_F^k x_F^l\mathcal{D}_{kl}\big[\eps\mathcal{H}(1+\eta)\partial^{-2}\partial_0\zeta - \eps\zeta + \eps\partial^{-2}\partial^2_0\zeta\big]\,\,, \label{eq:cfc_metric_general-components_riemann-explicit-slow_roll-1} \\ 
&h^F_{0i}(\tau_F,\vec{x}_F) = \mathcal{O}(\kl^3)\,\,, \label{eq:cfc_metric_general-components_riemann-explicit-slow_roll-2} \\
&h^F_{ij}(\tau_F,\vec{x}_F) = \bigg[x_F^k x_F^l\mathcal{D}_{kl}(\eps\mathcal{H}\partial^{-2}\partial_0\zeta) - \frac{K_F\abs{\vec{x}_F}^2}{2} \bigg]\delta_{ij}
\,\,, \label{eq:cfc_metric_general-components_riemann-explicit-slow_roll-3}
\end{align}
\end{subequations}
\ie a curved FLRW metric with $K_F\propto\partial^2\zeta$ and (slow-roll suppressed) tidal corrections. We can also write down the correction $(a_F/a)|_\ell$ to the scale factor, \ie \eq{a_F_over_a-A}, at linear 
order in perturbations (as we are doing throughout this section). We find 
\begin{equation}
\label{eq:a_F_over_a-B}
\begin{split}
\frac{a_F(P)}{a(P)} &= 1 + \int_{\tau_\ast}^{\tau_F}\dif s\,\bigg(\partial_0\zeta(s,\vec{0}) + \frac{1}{3}\partial_i V^i(s,\vec{0})\bigg) \\
&= 1 + \int_{\tau_\ast}^{\tau_F}\dif s\,\bigg(\partial_0\zeta(s,\vec{0}) - \frac{1}{3}\partial^2\psi(s,\vec{0})\bigg) + \mathcal{O}(\kl^4)\,\,.
\end{split}
\end{equation}
We will use this metric in \sect{field_theory}, where we will compute the full CFC bispectrum by working directly at the level of the action.

\section{Bispectrum transformation}
\label{sec:bispectrum_transformation}

\noindent In this section, we transform Maldacena's bispectrum to the conformal Fermi frame, following the approach of \cite{Pajer:2013ana}, to obtain our main result, \ie 
the bispectrum $B^F_\zeta(\vec{k}_s,\vec{k}_\ell)$. We split the computation into three steps:
\begin{itemize}[leftmargin=*]
\item in \sect{zeta_transformation} and \sect{ps_transformation}, respectively, we derive the transformation rules for the short-scale 
curvature perturbation and its power spectrum under the CFC change of coordinates; 
\item \sect{squeezed_bispectrum-A} contains the derivation of the bispectrum $B^F_\zeta(\vec{k}_s,\vec{k}_\ell)$ up to and including second order in gradients of the long mode.
\end{itemize}

\subsection{Transformation of the curvature perturbation}
\label{sec:zeta_transformation}

\noindent We start from the transformation of the 
curvature perturbation $\zeta$: we consider a coordinate transformation from $x$ to $\bx$ that does not change the hypersurfaces of constant $\tau$, \ie
\begin{subequations}
\label{eq:passive_change-A}
\begin{align}
&\tau=\tau(\bx) = \bar{\tau}\,\,, \label{eq:passive_change-A-1} \\
&x^i= x^i(\bx) = \bx^i + \xi^i(\bx)\,\,. \label{eq:passive_change-A-2}
\end{align}
\end{subequations}
Since $\tau = \bar{\tau}$, the metric on surfaces on constant time will now be given by 
\begin{equation}
\label{eq:passive_change-B}
\begin{split}
\bar{g}_{ij}(\bx) 
&= g_{ij}(x(\bx)) + g_{il}(x(\bx))\partial_j\xi^l(\bx) + g_{kj}(x(\bx))\partial_i\xi^k(\bx) + \mathcal{O}(\xi^2)\,\,,
\end{split}
\end{equation}
where derivatives are understood to be w.r.t. $\bar{x}$. 
Now, the curvature perturbation on surfaces of constant time is defined 
by \cite{Maldacena:2002vr,Rigopoulos:2003ak,Lyth:2004gb,Weinberg:2008nf,Weinberg:2008si,Assassi:2012et,Dias:2014msa}
\begin{equation}
\label{eq:zeta_delta_N_definition}
\bar{\zeta}(\bx) = \frac{\log\det(\bar{g}_{ij}(\bx)/a^2(\tau))}{6}\,\,,
\end{equation}
where $a$ is not changed since we are not transforming the time coordinate. 
If we work in the comoving gauge, we can write down $\bar{g}_{ij}(\bx)$ as (lowering spatial indices with $\delta_{ij}$) 
\begin{equation}
\label{eq:passive_change-C}
\begin{split}
&\bar{g}_{ij}(\bx)/a^2 = \delta_{ij} + \partial_i\xi_j(\bx) + \partial_j\xi_i(\bx) + (\underbrace{e^{2\zeta(x(\bx))}-1}_{\hphantom{\Delta g(\bx)\,} = \,\Delta g(\bx)})\delta_{ij} \\
&\hphantom{\bar{g}_{ij}(\bx)/a^2 =} + 
\Delta g(\bx)\big[\partial_i\xi_j(\bx) + \partial_j\xi_i(\bx)\big] + \mathcal{O}(\xi^2)\,\,. 
\end{split}
\end{equation}
Dropping terms cubic in perturbations (which we denote by ``$\dots$'' below), we arrive at
\begin{equation}
\label{eq:passive_change-D}
\begin{split}
&\log(\bar{g}_{ij}(\bx)/a^2) = \partial_i\xi_j(\bx) + \partial_j\xi_i(\bx) + \Delta g(\bx)\delta_{ij} + \Delta g(\bx)\big[\partial_i\xi_j(\bx) + \partial_j\xi_i(\bx)\big] \\
&\hphantom{\log(\bar{g}_{ij}(\bx)/a^2) =} -\frac{1}{2}\big[\Delta g(\bx)\big]^2\delta_{ij} - \Delta g(\bx)\big[\partial_i\xi_j(\bx) + \partial_j\xi_i(\bx)\big] + \dots \\
&\hphantom{\log(\bar{g}_{ij}(\bx)/a^2)} = \partial_i\xi_j(\bx) + \partial_j\xi_i(\bx) + 2\zeta(x(\bx))\delta_{ij}+\dots\,\,.
\end{split}
\end{equation}
Taking the trace of the above equation, we find 
\begin{equation}
\label{eq:passive_change-E}
\bar{\zeta}(\bx) = \frac{\partial_i\xi^i(\bx)}{3} + \zeta(x(\bx))\,\,.
\end{equation}
Now, we are interested in long-wavelength transformations, \ie $\xi^\mu = \xi^\mu_\ell$ will contain only long modes. Therefore, if we 
split also $\bar{\zeta}$ in long and short modes, we find that its short-scale part transforms as a scalar: $\bar{\zeta}_s(\bx) = \zetas(x(\bx))$.

This derivation does not hold if we change also the time coordinate. If one is interested in working at zeroth and linear order in gradients, as it 
was done in \cite{Pajer:2013ana}, this is not a problem since the change to CFC affects $\tau$ only at order $\kl^2$. 
However, for our purposes we will need to consider also the fact that surfaces of constant conformal time are not surfaces of constant 
CFC time. In \sect{appendix_active} we show that in this case the transformation rule for $\zetas$ is nontrivial, namely it acquires a shift 
\begin{equation}
\label{eq:passive_change-F}
\bar{\zeta}_s(\bx) = \zetas(x(\bx)) 
+ \frac{N^i_s(\bx)\partial_i\xi^0_\ell(\bx)}{3}\,\,.
\end{equation}
Since $N_i = \partial_i\psi$, with $\psi$ a function of $\zeta$, this additional shift will generate other terms proportional to (spatial derivatives) of the short-scale power 
spectrum $\braket{\zetas\zetas}$.

\subsection{Short-scale power spectrum transformation}
\label{sec:ps_transformation}

\noindent We 
can now see how the short-scale power spectrum $\braket{\zetas\zetas}$ of the curvature perturbation $\zeta$ is transformed when moving to the CFC frame. 
The 
overall transformation of $\braket{\zetas\zetas}$ will follow closely the one presented in \cite{Pajer:2013ana}, the main difference 
being the fact that 
we will go up to order 
$\kl^2$ in the gradient expansion. This implies that, in principle, we would need to take the transformation of conformal time (\ie the contribution of $\xi^0$) into 
account. However, it is straightforward to see that these terms will not matter on super-Hubble scales:
\begin{itemize}[leftmargin=*]
\item the first contribution is 
\begin{equation}
\label{eq:decay_superhorizon-A}
\zetas^F(x_F)\supset\frac{N^i_s(\fx)\partial_i\xi^0_\ell(\fx)}{3}\,\,.
\end{equation}
Since $N^i_s\sim-\partial_i(\zetas/\cH)$ and $\partial_i\xi^0_\ell$ go to zero for $-\ks\tau_F\ll 1$, $-\kl\tau_F\ll 1$, these terms in the transformation 
of the short-scale curvature perturbation can be dropped; 
\item the second contribution is, instead, given by
\begin{equation}
\label{eq:decay_superhorizon-B}
\zetas^F(\fx)\supset\xi^0_\ell(\fx)\partial_0\zetas(\fx)\,\,.
\end{equation}
Since $\zetas$ freezes on super-Hubble scales, we see that also this part of the transformation will not be relevant for $B_\zeta^F(\vec{k}_s,\vec{k}_\ell)$.
\end{itemize}
Then, for $\tau_F\to 0^-$, we can write the equal-time power spectrum of short modes in CFC as
\begin{equation}
\label{eq:ps_transformation-A}
\begin{split}
&\braket{\zetas^F(\fbx_1)\zetas^F(\fbx_2)} = \braket{\zetas(\vec{x}_1)\zetas(\vec{x}_2)}\,\,,
\end{split}
\end{equation}
where we have defined $x^i_{1,2}\equiv x^i(\vec{x}^F_{1,2})$, 
and we have dropped all time dependences for simplicity of notation. Now, thanks to translation invariance, we can write the short-scale power spectrum in real space as 
\begin{equation}
\label{eq:ps_transformation-B}
\braket{\zetas(\vec{x}_1)\zetas(\vec{x}_2)} = \braket{\zetas\zetas}(r)\,\,,
\end{equation}
where $\R\equiv\vec{x}_1 - \vec{x}_2$, and $r\equiv\abs{\R}$. 
We can now expand this at $\mathcal{O}[(r_F^i)^3]$ and to first order in long-wavelength perturbations: it is straightforward to see that
\begin{equation}
\label{eq:x1_minus_x2}
\begin{split}
&r^l_F = r^l_F + A^l_i(\fbx_c)\,r_F^i + \frac{1}{4}C^l_{kij}(\fbx_c)\,r_F^ir_F^jr_F^k\,\,,
\end{split}
\end{equation}
since we construct the CFC frame around 
$\fbx_c = (\fbx_1 + \fbx_2)/2$. The exact position of the central geodesic does not matter in the squeezed limit. This has been proven up to order $\kl$ 
in \cite{Pajer:2013ana}, and here we see that this is true also at order $\kl^2$: indeed, choosing the middle 
point gets rid of $B^l_{ij}(\fbx_c)$ only, which is of order $\kl$ (in fact, it is $\sim\delta_{ij}\partial^l\zetal(\fbx_c)$), 
and no terms of order $\kl^2$ are cancelled. That is, any additional correction to our result coming from the change in the position of 
the central geodesic enters at order $\kl^3$. The final expression for the power spectrum of the short modes in CFC, then, is given by
\begin{equation}
\label{eq:ps_transformation-space}
\begin{split}
&\braket{\zetas^F\zetas^F}(r_F) = \braket{\zetas\zetas}(r_F) +A^l_i(\fbx_c)\,r_F^i\partial_l
\braket{\zetas\zetas}(r_F) \\
&\hphantom{\braket{\zetas^F\zetas^F}(r_F) = } +\frac{1}{4}C^l_{kij}(\fbx_c)\,r_F^ir_F^jr_F^k\partial_l
\braket{\zetas\zetas}(r_F)\,\,.
\end{split}
\end{equation}

\subsection{Squeezed limit bispectrum in CFC -- first method}
\label{sec:squeezed_bispectrum-A}

\noindent The Maldacena consistency relation \cite{Maldacena:2002vr,Creminelli:2004yq,Creminelli:2011rh,Creminelli:2012ed,Hinterbichler:2013dpa} in global coordinates, \ie 
\begin{equation}
\label{eq:CS_squeezed-A}
B_{\zeta}(\vec{k}_s,\vec{k}_\ell) = -(\ns-1) P_{\zeta}(\ks)P_{\zeta}(\kl) + \mathcal{O}\bigg(\frac{\kl^2}{\ks^2}\bigg)\,\,,
\end{equation}
is equivalent to saying that a long-wavelength mode modulates the small-scale power 
as 
\begin{equation}
\label{eq:CS_squeezed-B}
P_{\zeta}(\ks)|_{\zeta(\kl)}
= [1 - (\ns-1) \zeta(\kl)] P_{\zeta}(\ks)\,\,.
\end{equation}
The transformation to CFC, up to linear order in $\kl/\ks$, cancels exactly the term $\propto(\ns-1)$ in the previous equation. 
We want to see, now, what are the terms that survive if we carry the CFC construction up to order $\kl^2/\ks^2$. 
Schematically, working in real space, we can write the transformation to CFC of the short-scale power spectrum as (we drop all ``$F$''s on coordinates for simplicity of notation) 
\begin{equation}
\label{eq:CS_squeezed-C}
\braket{\zetas\zetas}(r)|_{\zetal(\vec{x}_c)}\to \braket{\zetas\zetas}(r)|_{\zetal(\vec{x}_c)} + \Xi(\zetal(\vec{x}_c))\braket{\zetas\zetas}(r)\,\,,
\end{equation}
where $\Xi$ stands for the various terms, including derivatives w.r.t. $\vec{r}$, of \eq{ps_transformation-space}. 
If we multiply the r.h.s. of the above equation by $\zetal(\vec{x}_3)$, and then average over long modes, we can see what part of the 
long-short coupling is cancelled when we move to the CFC frame. 
Following \cite{Pajer:2013ana}, we can compute what is the contribution of these terms when we go in Fourier space 
$\x_1\leftrightarrow\vec{k}_1$, $\x_2\leftrightarrow \vec{k}_2$ and $\x_3\leftrightarrow\vec{k}_\ell$: 
\begin{itemize}[leftmargin=*]
\item the first term on the r.h.s. of 
the above equation will give the single-field slow-roll bispectrum in global coordinates of \cite{Maldacena:2002vr}, \ie 
\begin{equation}
\label{eq:maldacena_bispectrum-squeezed}
\begin{split}
B_\zeta(\vec{k}_s,\vec{k}_\ell) = P_{\zeta}(\ks)P_{\zeta}(\kl)\bigg\{(1-\ns) + \frac{\kl^2}{\ks^2}\bigg[&\bigg(\frac{29}{6}\eps + \frac{1}{4}\eta\bigg) \\
&- \bigg(\frac{1}{12}\eps + \frac{5}{8}\eta\bigg)(1-3\COS^2)\bigg]\bigg\}\,\,,
\end{split}
\end{equation}
where we have split the part $\propto\kl^2/\ks^2$ into a monopole and a quadrupole part. This shows how, for an isotropic long 
mode, the contribution of $\eps$ to the bispectrum of Maldacena at order $\kl^2/\ks^2$ is $\approx 20$ times larger than the one proportional to $\eta$; 
\item in \cite{Pajer:2013ana} it is shown how, thanks to translational invariance, the term coming from the coordinate transformation can be written in Fourier space as 
\begin{equation}
\label{eq:fourier_space_transformation}
\braket{\zetal(\vec{x}_3)\Xi(\zetal(\vec{x}_c))}\braket{\zetas\zetas}(r)\to\underbrace{P_{\zeta\Xi}(\kl)P_{\zeta}(\ks)|_{\vec{k}_s = \vec{k}_1 + \vec{k}_\ell/2}}_{\hphantom{\Delta B_\zeta(\vec{k}_s,\vec{k}_\ell)\,}\equiv\,\Delta B_\zeta(\vec{k}_s,\vec{k}_\ell)}\,\,,
\end{equation}
where we have omitted an overall $(2\pi)^3\delta(\vec{k}_1+\vec{k}_2+\vec{k}_\ell)$ of momentum conservation. 
\end{itemize}

This result allows us to compute separately the long- and short-wavelength power spectra. More precisely, when we go to Fourier space, we 
include directly in $\braket{\zetas\zetas}(r)$ the powers of $r^i$ and derivatives $\partial_i$ contained in $\Xi$. The full calculation is carried 
out in \sect{bispectrum_fourier}; here we cite the only result that we are going to need, that is 
\begin{equation}
\label{eq:trial_fourier_short}
(r^ir^jr^k\dots)\partial_l\braket{\zetas\zetas}(r)\to i^{N+1}\frac{\partial^{N}}{\partial\ks^i\partial\ks^j\partial\ks^k\dots}\big[\ks^lP_{\zeta}(\ks)\big]\,\,. 
\end{equation}
In $\braket{\zetal(\vec{x}_3)\Xi(\zetal(\vec{x}_c))}\equiv P_{\zetal\Xi}(\abs{\vec{x}_3 - \vec{x}_c})$, now, we will only have contributions like 
\begin{equation}
\label{eq:trial_fourier_long-A}
P_{\zetal\Xi}(\abs{\vec{x}_3 - \vec{x}_c})\supset\braket{\zetal(\vec{x}_3)\partial_{ijk\dots}\zetal(\vec{x}_c)}\,\,,
\end{equation}
that in Fourier space will read as 
\begin{equation}
\label{eq:trial_fourier_long-B}
\braket{\zetal(\vec{x}_3)\partial_{ijk\dots}\zetal(\vec{x}_c)}\to\big[(-i\kl^i)(-i\kl^j)(-i\kl^k)\dots\big]P_{\zeta}(k_\ell)\,\,.
\end{equation}

The two contributions that we must consider are $A^l_i(\vec{x}_c)$ and $C^l_{kij}(\vec{x}_c)$. Before embarking on the calculation, we 
note that $\Delta B_\zeta$ will be of order $\ns-1$: in fact, since we are basically just changing the way in which we measure distance, we 
will have an effect only if the short-scale power spectrum is not scale invariant. This tells us three things:
\begin{itemize}[leftmargin=*]
\item we can use the de Sitter mode functions \cite{Maldacena:2002vr,Chen:2006nt,Cheung:2007sv,Baumann:2009ds}, \emph{i.e.}, dropping irrelevant phases, 
\begin{equation}
\label{eq:can_use_dS_modes-A}
\zeta(\tau,k) = \zeta(0,k)(1+ik\tau)e^{-ik\tau} = \sqrt{P_{\zeta}(k)}\,(1+ik\tau)e^{-ik\tau}\,\,,
\end{equation}
to compute $(a_F/a)|_\ell$, that will enter in $A^l_i(\vec{x}_c)$. This is analogous to what is done in Maldacena's calculation 
of the bispectrum in global coordinates: once the cubic Lagrangian for $\zeta$ is found to be of second order in slow-roll (the quadratic one 
being of first order), the in-in computation of the leading order contribution to the three-point function can be carried out using just the de Sitter modes;
\item we can drop the slow-roll suppressed part of the shift constraint when we compute $(a_F/a)|_\ell$. That is, we can 
take $\psi_\ell$ to be just $-\zetal/\cH$ and drop $\eps\partial^{-2}\partial_0\zetal$, when we use the expression for $(a_F/a)|_\ell$ given in \eq{a_F_over_a-B}; 
\item when we consider $C_{kij}^l(\vec{x}_c)$, we can drop the $\eps$-suppressed part of the stereographic projection, \ie the last three lines of \eq{A_lkij-main_text}: the only contribution that 
we need to consider is the isotropic one, which involves the curvature $K_F$. 
\end{itemize} 
However, we note that there will be no need of actually computing $(a_F/a)|_\ell$: in fact, from its definition of \sect{cfc_metric_inflation} and 
our choice of initial time for the definition of Conformal Fermi Coordinates, we have that 
\begin{equation} 
\label{eq:only_constant}
\begin{split}
(a_F/a)|_\ell &\xrightarrow{-\kl\tau\,\ll\,1} C_{a_F} + \mathcal{O}(\kl^3)\,\,.
\end{split}
\end{equation}
Since we take $C_{a_F}$ to be zero, we can forget about this contribution. Then,
\begin{itemize}[leftmargin=*]
\item we start from $A^l_i(\vec{x}_c)$, which gives 
\begin{equation}
\label{eq:spatial_bispectrum-A_l_i-A}
\Delta B_\zeta^{(1)}(\vec{k}_s,\vec{k}_\ell) = P_{\zeta}(\kl)\prt{}{\ks^i}[\ks^iP_{\zeta}(
\ks)]\,\,,
\end{equation}
where 
\begin{equation}
\label{eq:can_use_dS_modes-B}
\prt{}{\ks^i}[\ks^iP_{\zeta}(
\ks)] = \bigg(3 + \frac{\dif}{\dif\log\ks}\bigg)P_{\zeta}(
\ks) = (\ns-1)P_{\zeta}(
\ks)\,\,.
\end{equation}
So we have
\begin{equation}
\label{eq:spatial_bispectrum-A_l_i-B}
\begin{split}
\Delta B_\zeta^{(1)}(\vec{k}_s,\vec{k}_\ell) &= (\ns-1) P_{\zeta}(
\kl
)
P_{\zeta}(
\ks)\,\,;
\end{split}
\end{equation}
\item the second (and last) term we have to consider is $C^l_{kij}(\vec{x}_c)$. It contains two contributions. One from $\sim\partial_i\partial_j\zeta$, and 
one from $\sim\cH\partial_i\partial_j\digamma$ (as we see from \tab{conformal_christoffels}): since $\digamma$ is already of order $\kl^2$, it is sufficient to include the former. At leading order in slow-roll, then, we have
\begin{equation}
\label{eq:spatial_bispectrum-C_l_kij-A}
\begin{split}
C^l_{kij}(\kl) = -\frac{1}{6}
\bigg[&\delta_{ij}\kl^k\kl^l\zeta(\kl) - \delta^l_j \kl^k\kl^i\zeta(\kl) - \delta^l_i\kl^k\kl^j\zeta(\kl) \\
&+\frac{1}{9}(\delta^l_k\delta_{ij} + \delta^l_i\delta_{jk}+\delta^l_j\delta_{ki})\kl^2\zeta(\kl)\bigg]\,\,.
\end{split}
\end{equation}
In the above equation, if we isolate a tensor $\mathcal{L}^l_{kij}\propto\kl^2\zeta(\kl)$, we can write 
\begin{equation}
\label{eq:spatial_bispectrum-C_l_kij-B}
\begin{split}
\Delta B_\zeta^{(2)}(\vec{k}_s,\vec{k}_\ell) &= 
\frac{1}{4}P_{\zeta}(
\kl
)\mathcal{L}_{kij}^l\bigg[i^{4}\frac{\partial^{3}}{\partial\ks^i\partial\ks^j\partial\ks^k}\big[\ks^lP_{\zeta}(
\ks)\big]\bigg] \\
&\equiv \frac{1}{4}P_{\zeta}(
\kl
)\mathcal{L}_{kij}^l\mathcal{S}^l_{ijk}P_{\zeta}(
\ks)\,\,.
\end{split}
\end{equation}
We compute this quantity in \sect{bispectrum_fourier} and cite here only the final result, \ie\footnote{This contribution vanishes for an isotropic 
long mode (\ie $\mu^2 = 1/3$). Indeed, in this case it is easy to see that to go from the metric in global coordinates to the CFC 
metric described in \sect{cfc_metric_inflation} it is enough to remove the constant and constant gradient parts of metric 
perturbations. In fact, $g_{0i}$ is already zero in the isotropic case and the curvature part of $g_{ij}$ (\ie the one proportional to $\abs{\vec{x}}^2$) is already of the right form.} 
\begin{equation}
\label{eq:spatial_bispectrum-C_l_kij-C}
\Delta B_\zeta^{(2)}(
\vec{k}_s,\vec{k}_\ell) = (\ns-1)\frac{\kl^2}{\ks^2}\bigg(-\frac{5}{24} + \frac{5}{8}\mu^2\bigg)P_\zeta(\kl)P_\zeta(\ks)\,\,.
\end{equation}
\end{itemize}
Summing these two contributions to the Maldacena bispectrum of \eq{maldacena_bispectrum-squeezed}, we see that in the 
CFC frame the long-short coupling still retains terms $\propto\eta$: more precisely, we have
\begin{equation}
\label{eq:cfc_bispectrum}
\begin{split}
B_\zeta^F(\vec{k}_s,\vec{k}_\ell) &= \frac{\kl^2}{\ks^2} P_{\zeta}(\ks)P_{\zeta}(\kl)\bigg[\bigg(\frac{29}{6}\eps + \frac{1}{4}\eta\bigg) + \bigg(\frac{1}{3}\eps - \frac{5}{12}\eta\bigg)(1-3\COS^2)\bigg] \\ 
&= \frac{\kl^2}{\ks^2} P_{\zeta}(\ks)P_{\zeta}(\kl)\bigg[\bigg(\frac{13}{3}\eps - \frac{1}{4}(\ns-1)\bigg) + \bigg(\frac{7}{6}\eps + \frac{5}{12}(\ns-1)\bigg)(1-3\COS^2)\bigg]\,\,,
\end{split}
\end{equation}
where, in the last line, we have highlighted the tilt of the scalar spectrum $\ns-1$ 
instead of $\eta$. 
We see that, as in Maldacena's squeezed bispectrum at order $\kl^2/\ks^2$, the contribution to the physical isotropic mode coupling 
$\propto\eps$ is larger than the one 
$\propto\eta$ by a factor of $\approx 20$.

For reference, we can match the result \eq{cfc_bispectrum} to the 
squeezed limit of the equilateral shape $\mathcal{S}_\mathrm{equil.}(k_1,k_2,k_3)$. Using Eq.~(52) from \cite{Schmidt:2010gw}, we have 
\begin{equation}
B_\zeta^\text{equil.}(\vec{k}_s,\vec{k}_\ell) = \frac{\kl^2}{\ks^2} P_{\zeta}(\ks)P_{\zeta}(\kl)\, 4f_{\mathrm{NL}}^\text{equil.}\left[2 + (1-3\mu^2) \right]\,.
\end{equation}
Clearly, \eq{cfc_bispectrum} cannot be matched to the equilateral shape, since the relative isotropic and anisotropic contributions are different. Moreover, we have only calculated the $\mathcal{O}(k_\ell^2/k_s^2)$ contribution in the squeezed limit, and different shapes $\mathcal{S}(k_1,k_2,k_3)$ can have the same squeezed-limit scaling $\propto k_\ell^2/k_s^2$. Hence, we caution against associating \eq{cfc_bispectrum} with the equilateral template. 
Roughly, however, \eq{cfc_bispectrum} corresponds to $f_{\mathrm{NL}}^\text{equil.} \sim 0.1\times (n_\mathrm{s}-1)$.

\section{Calculation at the level of the action}
\label{sec:field_theory}

\noindent In this section we derive the CFC bispectrum in the limit $\eps\to 0$ directly, without recurring to a transformation of Maldacena's result.

\subsection{Short modes in CFC}
\label{sec:short_modes_cfc}

\noindent We start by defining the short modes in CFC: keeping the ``$F$'' label on coordinates and components to 
parallel the first part of our previous calculation, we write the line element in CFC coordinates as
\begin{equation}
\label{eq:cfc_LS_line_element}
\begin{split}
&\dif s^2_F = -a^2\big[1 + 2 (N_1^F)_\ell + 2(N_{1}^F)_s\big]\dif\tau^2_F \\
&\hphantom{\dif s^2_F =} + a^2
(N_i^F)_s
(\dif\tau_F\dif x^i_F + \dif x^i_F\dif\tau_F) \\
&\hphantom{\dif s^2_F =} + a^2e^{2\zetal^F}e^{2\zetas^F}\delta_{ij}\dif x^i_F\dif x^j_F\,\,,
\end{split}
\end{equation}
where:
\begin{itemize}[leftmargin=*]
\item we have put to zero $(N_i^F)_\ell$, since we have seen that the time-space components of the long-wavelength metric in CFC are of order $\kl^3$;
\item working at linear order in the long mode (as we are doing throughout the paper), the long-wavelength 
part of the metric can be related, by direct comparison, to the results of \sect{cfc}. For example, the anisotropic part of $\zetal^F$ will be 
\begin{equation}
\label{eq:anisotropic_zeta_L}
(\zetal^F)(x_F)^\mathrm{anis.} = \frac{1}{2}x^i_F x^j_F\mathcal{D}_{ij}\big[\eps\mathcal{H}\partial^{-2}\partial_0\zetal(\tau,\vec{0})\big]\,\,,
\end{equation}
where $\zetal$ is the long-wavelength curvature perturbation in global coordinates; 
\item we have included the modification to the scale factor, \ie 
\begin{equation}
\label{eq:uniform_zeta_L}
a_F(\tau_F) = a(\tau_F)\big[1 + (a_F/a)(\tau_F)|_\ell + \cH\xi^0(\tau_F,\vec{0})\big]\,\,,
\end{equation}
directly into $\zetal^F$ and $(N_1^F)_\ell$: by doing so we can keep track more easily of both the order in perturbations and the order in the slow-roll expansion; 
\item $\zetas^F$, $(N_1^F)_s$ and $(N_i^F)_s = \partial_i\psi_s^F$ (whose indices will be raised and 
lowered with $\delta_i^j$) are the short modes. As before, we stopped at first order in perturbations in the small-scale lapse and shift constraints, which will be solved linearly in terms of $\zetas$. 
\end{itemize}
At this point, one can write down the action for $\zetas^F$: the lapse and shift constraints will have the usual expression, and the quadratic action $S_{(2)}$ will be given by \cite{Maldacena:2002vr} 
\begin{equation}
\label{eq:quadratic_action_zeta_S}
S_{(2)} = \int\dif^4x_F\,a^2\eps\big[(\partial_0\zetas^F)^2 - (\partial_i\zetas^F)^2\big]\,\,. 
\end{equation}
Then, the goal is to compute the power spectrum of $\zetas^F$ in the background of the CFC long-wavelength metric: since the latter is 
explicitly of order $\kl^2$, it is clear that the CFC bispectrum will vanish at zeroth and first order in gradients of the long mode. Now, in order to calculate 
the $\mathcal{O}(\kl^2)$ contribution, one needs the cubic action with one long leg and two short legs. This can be computed with the standard 
methods (see \cite{Maldacena:2002vr,Chen:2006nt}, for example) and, as in the standard case, the ``brute-force'' computation gives 
an action which is of order zero in slow-roll (compared with the quadratic 
action of \eq{quadratic_action_zeta_S}, which is slow-roll suppressed): however, due to the complicated relation between $\zetal^F$ and $(N_1^F)_\ell$,\footnote{This relation can 
be found by solving at linear order the lapse constraint for $(N_1^F)_\ell$ with the metric of \eq{cfc_LS_line_element}. The result is not particularly illuminating, so we 
will not write it down.} integrating by parts to remain with $\mathcal{O}(\eps)$ terms that can be removed with a field redefinition is more difficult than in Maldacena's calculation. 

A possible alternative approach is to work in $\phi$ gauge for the short modes, while keeping the long-wavelength metric as in \eq{cfc_LS_line_element}. In this case, the 
quadratic action for the small-scale field fluctuations $\vphi_s$ would be of order zero in slow-roll \cite{Maldacena:2002vr,Chen:2006nt}, but this would still not 
help because the cubic action will again, naively, not be slow-roll suppressed (we refer also to \sect{small_sound_speed} for a more detailed discussion about 
these issues). In both these cases, then, we could not do an in-in calculation using the de Sitter modes $\propto e^{-ik\tau}(1+ik\tau)$, since we would be 
missing terms due to slow-roll corrections to the mode functions: we would need to use the full solution of the classical equations of motion for the 
short modes in terms of Hankel functions, complicating the in-in integral considerably. For this reason, we will employ a different method, that is explained in the following section.

\subsection{From flat gauge to CFC}
\label{sec:from_flat_to_cfc}

\noindent This second method is based on the observation that, in flat gauge, all interactions are suppressed by $\sqrt{\eps}$. Therefore, if we are 
interested only in the contribution to the CFC bispectrum proportional to $\eta$ (which is the focus of this work), we expect that it will not be necessary 
to do any in-in calculation. We will explain how this comes about in the following. 

To simplify notation, in this section we will use $x=(x^0,\vec{x})$ for global coordinates, $x_F=(x^0_F,\vec{x}_F)$ for CFC. 
We use a ``tilde'' and a ``hat'' for coordinate changes, and a ``prime'' for time derivatives of the background inflaton $\bar{\phi}$). 
We then proceed in the following way: 
\begin{itemize}[leftmargin=*]
\item we start from the long-wavelength metric in global coordinates $x$, in $\zeta$ gauge. At linear order in the long mode, we can 
go to flat gauge 
with a simple time shift, \ie
\begin{equation}
\label{eq:flat_cfc-i}
x^0 = \tilde{x}^0 - \frac{\zetal}{\cH}\,\,.
\end{equation}
This coordinate change will originate an inflaton perturbation $\vphi_\ell = -\sqrt{2\eps}\,\zetal$, which is of order zero in slow-roll since a factor of $1/\sqrt{\eps}$ is ``hidden'' in 
$\zetal$; 
\item to this $\vphi_\ell$ we add a short $\vphi_s$, solve the constraints, and compute both the quadratic action for $\vphi_s$ and the interaction 
terms at cubic order (see also \cite{Pajer:2016ieg} for details):
\begin{itemize}
\item $S_{(2)}$ is given by 
\begin{equation}
\label{eq:flat_cfc-ii}
S_{(2)} = \int\dif^4\tilde{x}\,a^2\big[(\partial_0\vphi_s)^2 - (\partial_i\vphi_s)^2 + \cH^2\eta\,\vphi_s^2\big]\,\,.
\end{equation}
So, we see that $\eta$ provides a mass for $\vphi_s$, which tells us that $\vphi_s$ will not be conserved on super-Hubble scales; 
\item the result for the cubic terms will be Maldacena's cubic Lagrangian in flat gauge, with one long leg and two short legs. It is then easy to see 
from Eq.~(3.8) of \cite{Maldacena:2002vr} that, at leading order in slow-roll, interactions will be suppressed by a factor $\sqrt{\eps}$: therefore in the 
limit $\eps\to 0$ there is no coupling between the long mode and short modes, \ie we can schematically write 
\begin{equation}
\label{eq:flat_cfc-iii}
P_{\vphi_s}|_{\vphi_\ell} = P_{\vphi_s}\,\,,
\end{equation}
where $P_{\vphi_s}$ is the usual power spectrum of $\vphi_s$ in an unperturbed FLRW background that one computes from \eq{flat_cfc-ii}, while $P_{\vphi_s}|_{\vphi_\ell}$ is 
the power spectrum of $\vphi_s$ in the background of a long-wavelength mode (\ie considering the coupling with $\vphi_\ell$); 
\end{itemize}
\item then, we transform from the flat gauge in global coordinates to CFC. At linear order, the transformation is simply given by 
\begin{equation}
\label{eq:flat_cfc-iv}
\tilde{x}^\mu = x_F^\mu + \underbrace{\frac{\zetal}{\cH}\delta^\mu_0 + \xi^\mu_\ell}_{\hphantom{\Delta^\mu_\ell\,}\equiv\,\Delta^\mu_\ell}\,\,,
\end{equation}
where $\xi^\mu_\ell$ is the vector field given in \eqsI{ABC_definition} in terms of $\zetal$. Now, after this change of coordinates, the spatial part 
of the metric at quadratic order in perturbations but at linear order in the long mode,\footnote{That is, we consider only 
quadratic terms that mix long and short modes: therefore, we do not consider the long part of $\tilde{g}_{0i}$ (which will not 
be of order $\kl^3$ yet) since it will give rise to terms quadratic in the long mode in the transformed spatial metric.} will be given by 
\begin{equation}
\label{eq:flat_cfc-v}
\begin{split}
g_{ij}^F &= -a^2\prt{\Delta^0_\ell}{x^i_F}\prt{\Delta^0_\ell}{x^j_F} + \prt{\Delta^0_\ell}{x^i_F}\tilde{g}_{0j} + \prt{\Delta^0_\ell}{x^j_F}\tilde{g}_{i0} + a^2e^{2\zetal^F}\delta_{ij} \\
&= a^2\prt{\Delta^0_\ell}{x^i_F}\partial_j\tilde{\psi}_s + a^2\prt{\Delta^0_\ell}{x^j_F}\partial_i\tilde{\psi}_s + a^2e^{2\zetal^F}\delta_{ij} + \mathcal{O}[(\zetal)^2]\,\,,
\end{split}
\end{equation}
where $\tilde{\psi}_s$ is the short-scale shift constraint in flat gauge, \ie $-\eps\partial^{-2}\partial_0(\cH\vphi_s/\bar{\phi}')$, and we 
dropped terms quadratic in $\zetal$. Correspondingly, the inflaton will transform as 
\begin{equation}
\label{eq:flat_cfc-vi}
\begin{split}
&\phi = \bar{\phi} + \sqrt{2\eps}\,\cH\xi^0_\ell + \vphi_s + \Delta^\mu_\ell\partial_\mu\vphi_s\,\,,
\end{split}
\end{equation}
where we used $\bar{\phi}' = \sqrt{2\eps}\,\cH$; 
\item then: we want to find the relation between $\vphi_s$ and $\zetas^F$, defined as in \eq{cfc_LS_line_element}. In order to do this, we 
first do a time translation $x^0_F = \hat{x}^0_F + T$ (with $T$ starting linear in short modes, and having a quadratic long-short coupling) that 
brings $\phi$ to $\bar{\phi} + \sqrt{2\eps}\,\cH\xi^0_\ell$. It is easy to see that $T$ is given by 
\begin{equation}
\label{eq:flat_cfc-vii}
\begin{split}
T&=-\frac{1}{\sqrt{2\eps}\,\cH}\bigg[\vphi_s + \Delta^\mu_\ell\partial_\mu\vphi_s - \frac{\bar{\phi}''}{\bar{\phi}'}\xi^0_\ell\vphi_s - \partial_0\xi^0_\ell\vphi_s\bigg] \\
&=-\frac{1}{\sqrt{2\eps}\,\cH}\bigg[\vphi_s + \Delta^\mu_\ell\partial_\mu\vphi_s - \cH\bigg(1-\eps+\frac{\eta}{2}\bigg)\xi^0_\ell\vphi_s - \partial_0\xi^0_\ell\vphi_s\bigg]\,\,.
\end{split}
\end{equation}
Now, we focus on the spatial metric $\hat{g}_{ij}$ after this time translation, working at quadratic order in perturbations but dropping terms involving two short modes. It will be given by 
\begin{equation}
\label{eq:flat_cfc-viii}
\hat{g}^F_{ij} = a^2\prt{\Delta^0_\ell}{\hat{x}^i_F}\partial_j\tilde{\psi}_s 
+ a^2\prt{\Delta^0_\ell}{\hat{x}_F^j}\partial_i\tilde{\psi}_s 
+ a^2e^{\zetal^F}e^{\zetas^F}\delta_{ij}\,\,,
\end{equation}
where we have used the fact that the long-wavelength part of $g_{0i}^F$ is 
$\mathcal{O}(\kl^3)$ (so that we can safely neglect its contribution to the transformation at the order we are working at), and we have defined $\zetas$ as 
\begin{equation}
\label{eq:flat_cfc-ix}
\zetas^F = \cH T + T \partial_0\zetal = \cH T - \frac{\vphi_s}{\sqrt{2\eps}}\partial_0\zetal\,\,.
\end{equation}
With some hindsight, then, we can also define $\zetas$ as
\begin{equation}
\label{eq:flat_cfc-x}
\zetas = -\frac{\cH}{\bar{\phi}'}\vphi_s = -\frac{\vphi_s}{\sqrt{2\eps}}\,\,,
\end{equation}
so that $\zetas^F$ becomes 
\begin{equation}
\label{eq:flat_cfc-xi}
\zetas^F = \cH T + 
\zetas\partial_0\zetal\,\,.
\end{equation}
We note that $\hat{g}^F_{ij}$ is not yet of the form of \eq{cfc_LS_line_element} because of the terms in \eq{flat_cfc-viii} involving the short-scale shift constraint in flat gauge, which is given by 
\begin{equation}
\label{eq:flat_cfc-xii}
\tilde{\psi}_s = -\eps\partial^{-2}\partial_0\bigg(\frac{\cH}{\bar{\phi}'}\vphi_s\bigg) = \eps\partial^{-2}\partial_0\zetas\,\,.
\end{equation}
However, we can see that these terms will not matter on super-Hubble scales. We can follow the approach of Maldacena: with a second 
order long-short spatial transformation (which does not modify the field perturbations at the order we are working at) we can remove these 
terms at the price of new second order contributions to $\zetas^F$. From \cite{Maldacena:2002vr} we can see that all the new terms 
that $\zetas^F$ gains will contain $\tilde{\psi}_s$, that is proportional to $\partial^{-2}\partial_0 \zetas$. However, we know that $\zetas$ must freeze 
on super-Hubble scales (this can be seen also at the level of the quadratic action, that can be derived from the action for $\vphi_s$ with the changes 
of coordinates discussed above). This tells us that we can safely neglect the contributions from this second order spatial transformations in the relation \mbox{between $\vphi_s$ and $\zetas^F$;}
\item then, we can focus just on \eq{flat_cfc-x}. We consider only terms that are either of order zero in slow-roll, or suppressed by $\eta$, dropping all 
terms $\propto\eps$. With these assumptions, $\zetas^F$ becomes equal to 
\begin{equation}
\label{eq:flat_cfc-xiii}
\begin{split}
&\zetas^F = \zetas + \xi^i_\ell\partial_i \zetas + \xi^0_\ell\partial_0 \zetas + \frac{1}{\cH}\zetal\partial_0 
\zetas \\
&\hphantom{\zetas^F =} + \frac{\eta}{2}\zetal\zetas - \zetas \partial_0\xi^0_\ell - \cH\xi^0_\ell \zetas + \zetas\partial_0\zetal^F \,\,.
\end{split}
\end{equation}
In the above equation we recognize the term $\xi^i_\ell\partial_i\zetas$ from \sect{squeezed_bispectrum-A}. We also see that both terms 
containing $\partial_0 \zetas$ will not contribute on super-Hubble scales, so they can be dropped. We deal with the remaining terms 
separately by considering that $\xi^0_\ell$ and $\zetal$ can be split in a uniform (which encodes the modified expansion history), isotropic and anisotropic part: 
\begin{itemize}
\item we start from the isotropic part. For $\xi^0_\ell$ it is zero, while for $\partial_0\zetal^F$ it is proportional to $\abs{\vec{x}_F}^2\partial_0 K_F$, which in turn is $\propto\kl^4$; 
\item the uniform part of $\partial_0\zetal^F$ is, dropping $\eps$-suppressed terms, equal to $\partial_0\xi^0_\ell + \cH\xi^0_\ell + \partial_0(a_F/a)|_\ell$. The 
first two terms exactly cancel with those in \eq{flat_cfc-xiii}, while from the definition of $(a_F/a)|_\ell$ discussed in \sect{cfc_metric_inflation} we see that 
the last one vanishes on super-Hubble scales;
\item finally, we can easily see from the results of \sect{cfc_metric_inflation} that the anisotropic part of $\partial_0\zetal^F$ is of order $\eps$ (or higher), while 
that of $\partial_0\xi^0_\ell+\cH\xi^0_\ell$ contains either $\eps$-suppressed terms, or terms that go to zero as fast as $\cH^{-2}$. 
\end{itemize}
So we conclude that the only relevant terms in \eq{flat_cfc-xi} will be 
\begin{equation}
\label{eq:flat_cfc-xiv}
\zetas^F = \zetas + \xi^i_\ell\partial_i \zetas + \frac{\eta}{2}\zetal\zetas\,\,.
\end{equation}
\end{itemize}

\subsection{Squeezed limit bispectrum in CFC -- second method}
\label{sec:squeezed_bispectrum-B}

\noindent We are now in a position to compute the squeezed limit bispectrum in the conformal Fermi frame. Since we 
have the power spectrum of $\vphi_s$, we can compute the power spectrum of $\zetas^F$ in the background of the long 
modes. Schematically, since $\zetas^F$ is $\zetas$ plus a long-short coupling, we would have 
\begin{equation}
\label{eq:flat_cfc-xv}
\braket{\zetas^F\zetas^F}|_{\zetal} = \braket{\zetas\zetas}|_{\zetal} + \mathcal{O}(\zetal)\braket{\zetas\zetas}\,\,.
\end{equation}
Had we kept also $\eps$-suppressed interactions in our flat gauge calculation of \sect{from_flat_to_cfc}, the first term on the r.h.s. of the 
above equation would actually also contain a coupling with long modes: however, we do not care about this term (since we are trying to capture only the 
part of the bispectrum proportional to $\eta$). Then, $\braket{\zetas\zetas}|_{\zetal}$ will just be the power spectrum of $\zetas$ computed from the 
quadratic Lagrangian of $\vphi_s$, \ie what we called $\braket{\zetas\zetas}$ in the previous section. The second term on the r.h.s. of \eq{flat_cfc-xv} contains 
both the contribution of $\xi^i_\ell\partial_i\zetas$, which reproduces exactly what we have computed in \sect{squeezed_bispectrum-A}, and a second term $\propto\eta$. We can 
deal with the latter by expanding $\zetal$ in a Taylor series around $\vec{x}^F_c\equiv(\vec{x}^F_1+\vec{x}^F_2)/2$, so 
\begin{equation}
\label{eq:flat_cfc-xvi}
\braket{\zetas^F(\vec{x}^F_1)\zetas^F(\vec{x}^F_2)}|_{\zetal(\vec{x}^F_c)}\supset\frac{\eta}{2}\bigg[2\zetal (\vec{x}^F_c) + \frac{r^i_Fr^j_F}{4}\partial_i\partial_j\zetal (\vec{x}^F_c)\bigg]\braket{\zetas\zetas}(r_F) + \mathcal{O}(\kl^3)\,\,.
\end{equation}
Going to Fourier space using the results of \sect{bispectrum_fourier}, more precisely the fact that 
\begin{equation}
\label{eq:flat_cfc-xvii}
\frac{\partial^2}{\partial\ks^i\partial\ks^j}P_{\zeta}(\ks) = -3\frac{\delta_{ij}}{\ks^2}P_{\zeta}(\ks) + 15\frac{\ks^i\ks^j}{\ks^4}P_{\zeta}(\ks) + \mathcal{O}(\ns-1)\,\,,
\end{equation}
and averaging over the long-wavelength perturbations, we reproduce the $\eta$ part of Maldacena's bispectrum in the squeezed limit, up to 
and including $\mathcal{O}(\kl^2/\ks^2)$. Summing this to the other contribution (noting that the first term in \eq{flat_cfc-xv} will not matter once we average 
over long modes, since it has no coupling to them that are proportional to $\eta$), we reproduce our final result of \eq{cfc_bispectrum} for $\eps\to 0$. This concludes 
our analysis: we stress that this method is not completely independent from that of \sect{bispectrum_transformation}, since we still need to compute what is 
the effect of the shift $\xi^i_\ell\partial_i\zetas$, but we consider it different enough to provide a consistency check.

\section{Interactions during inflation}
\label{sec:argument}

\noindent In this section we discuss an argument, put forward in \cite{Alvarez:2014vva}, to estimate the size of (gravitational) interactions between 
long and short modes. More generally, we review how the contribution $\propto\eta$ arises in Maldacena's bispectrum, 
and argue that $\eta$ must be locally observable, as shown through the direct calculation in \sect{bispectrum_transformation} and \sect{field_theory}.

\subsection{Where does \texorpdfstring{$\eta$}{\textbackslash eta} come from?}
\label{sec:eta}

\noindent Let us start by considering short-scale scalar field perturbations $\vphi_s$ in the separate universe (similarly to the setup described in \fig{figure_region}). Naively, one 
might think that a coupling to $\zetal$ enters at order $\eps^0$ \cite{Baldauf:2011bh,Creminelli:2013cga}: 
for example, the Ricci three-scalar on constant time hypersurfaces, which measures the spatial curvature, is $^{(3)}R\propto\partial^2\zetal$; that is, it is not 
slow-roll suppressed. Indeed, if one were to do a brute-force computation of the action for $\vphi_s$ in the long-wavelength background modified by $\zetal$ (\ie the 
cubic action with two short legs $\vphi_s$ and one long leg $\zetal$, which controls the interactions between the long and short modes), the result would naively 
appear to be of such order. However, one can compute the full spacetime Riemann tensor $R_{\mu\nu\rho\sigma}$ of the background (in any gauge\footnote{This expression is 
covariant, but not \textit{manifestly} covariant because we are trying to make explicit the dependence on $\eps$ and $\zetal$, which are defined in global FLRW coordinates.})
\begin{equation}
\label{eq:riemann_dS}
R_{\mu\nu\rho\sigma} = H^2(g_{\mu\rho}g_{\nu\sigma} - g_{\mu\sigma}g_{\nu\rho}) + \mathcal{O}(\eps\times\partial^2\zetal)\,\,,
\end{equation}
where we stopped at quadratic order in gradients of $\zetal$, and used the fact 
that time derivatives of $\zetal$ are also 
$\propto\partial^2\zetal$. In a general set of coordinates, $g_{\mu\nu}$ might contain unsuppressed terms of order $\eps^{0}\zetal$. We notice though 
that the leading term is the Riemann tensor for a maximally symmetric spacetime with Ricci 
scalar $\propto H^2$, namely de Sitter spacetime. Therefore, up to terms of order $\eps$, 
there must exist a change of coordinates that removes \emph{completely} the long mode from the right hand side.\footnote{In passing, we also note 
that this is the reason why in \sect{cfc_metric_inflation} we have seen that the anisotropic part of the long-wavelength metric in CFC is slow-roll 
suppressed. Indeed, de Sitter is an isotropic spacetime.} 
Then, the coupling between $\vphi_s$ and $\zetal$ is suppressed by $\eps$, and no term $\propto\eta$ only appears. We also know that these small-scale inflaton fluctuations 
have non-zero mass. \eq{flat_cfc-ii} tells us that this mass is $\propto\eta$. Therefore, $\vphi_s$ evolves on super-Hubble scales. Switching from inflaton perturbations 
to curvature perturbations cancels this time dependence, and induces an additional term $\propto\eta$ in the long-short mode coupling, since the relation 
between $\vphi$ and $\zeta$ is non-linear. For this reason, we can regard $\eta$ as measuring a \emph{physical} effect, \ie the time evolution of inflaton 
correlators on super-Hubble scales, and we do not expect $\eta$ to 
vanish in the CFC bispectrum at order $\kl^2/\ks^2$. 

Another way to look at this is to work directly with short-scale curvature perturbations $\zetas$: as Maldacena has shown, a straightforward 
computation of the cubic action of $\zeta$ with $\zetal$ in one leg and $\zetas$ in the other two leads to $S_{(3)} \sim\eps^0\times\zetal\times\zetas^2$. 
However, one can do a sequence of integration by parts to rewrite this as $S_{(3)}\sim\eps(1+\eps+\eta)\times\zetal\times\zetas^2$, with 
the term proportional to $\eps\eta$ arising when one integrates by parts terms such as $a^2\eps\,\zeta(\partial_0\zeta)^2$. This shows that 
also $S_{(3)}$ goes to zero when $\eps$ goes to zero. However, what matters is 
the relative order in slow-roll between the quadratic action $S_{(2)} \sim\eps\times\zetas^2$ and this cubic action. 
The quadratic action for $\zetas$ is also suppressed by $\eps$, so the size of interactions is $\sim (1+\eps+\eta)$: $\eta$ and $\eps$ are \emph{both} a measure 
of the coupling between long- and short-wavelength modes of $\zeta$. 
The fact that the background spacetime is de Sitter in the $\eps\rightarrow 0$ limit, even in presence of $\zetal$, 
does \emph{not} allow us to conclude that such long-wavelength perturbations have no effect on the short modes $\zetas$.\footnote{The argument we 
made for scalar perturbations, using \eq{riemann_dS}, does not apply to curvature perturbations. In fact, $\zetas$ is not a (perturbation of a) scalar 
field: it is a component of the metric which is non-linearly related to the inflaton $\vphi_s$ and has a non-minimal coupling with the Riemann tensor 
of the long-wavelength background.} 
Notice that the terms of order $\eps\times 1$ in $S_{(3)}$ do not contribute to correlation functions on super-Hubble scales (their contribution 
in the in-in calculation of the bispectrum decays). In fact, we know that the final result for the $2$-point function of $\zetas$ in presence of $\zetal$ must satisfy 
the consistency relation in the squeezed limit\mbox{, \ie at leading order in derivatives of $\zetal$ we have} 
\begin{equation}
\label{eq:local_PS}
\braket{\zetas\zetas}(r)\Big|_{\zetal(\vec{x}_c)} = [1+(1-\ns)\zetal(\vec{x}_c)]\braket{\zetas\zetas}(r)
\,\,.
\end{equation}
Then, one can do a counting of factors of $\eps$ and $\eta$ to see what is the order in slow-roll of the above expression. The tilt contains both $\eps$ and $\eta$, while the three perturbations 
of $\zeta$ each contain $1/\sqrt{\eps}$ (recall that $\braket{\zeta\zeta}\sim H^2/\eps$). Therefore the overall order of \eq{local_PS} is $\sim(\eps,\eta)\times\eps^{-3/2}$. One can then 
repeat the same argument for the full in-in calculation of this position-dependent power spectrum. From two powers of the short modes, of which we compute the $2$-point function in presence of $\zetal$, we have $(1/\sqrt{\eps}\,)^2$, while $S_{(3)}$ would give
\begin{equation}
\label{eq:order_SR}
S_{(3)}\sim\eps(1+\eps+\eta)\times\zetal\times\zetas^2\sim\eps(1+\eps+\eta)\times\eps^{-3/2}\,\,.
\end{equation}
Overall, we have $(1+\eps+\eta)\times\eps^{-3/2}$: to agree with the result in the squeezed limit, then, the terms of 
order $\eps\times 1$ in $S_{(3)}$ must not contribute on super-Hubble scales. By continuity, the same applies to other momentum configurations away from the squeezed limit.

\subsection{Interactions from non-trivial speed of sound}
\label{sec:c_s}

\noindent We conclude this section 
by briefly discussing the case where the inflaton speed of sound $\cs$ is different from $1$. In this case, we know that the operator giving $\cs\neq 1$ also induces 
cubic couplings for the inflaton \cite{Cheung:2007st}, leading to enhanced non-Gaussianities. Indeed, while the three-point function from these inflaton self-interactions still 
satisfies the consistency relation in the squeezed limit, the term proportional to $\kl^2/\ks^2$ is of order $(1-\cs^2)/\cs^2$ \cite{Seery:2005wm,Chen:2006nt,Creminelli:2013cga}, 
which 
can be much larger than the one coming from gravitational interactions for $\cs<1$. It is then easy to see how this still holds in the conformal Fermi frame: the 
corrections to the bispectrum coming from the transformation to CFC are of order of the scale dependence of the power spectrum, namely
\begin{equation}
\label{eq:c_s-A}
\Delta B_\zeta(\vks,\vkl)\sim\frac{\kl^2}{\ks^2}\times\frac{\dif\log[\ks^3P_{\zeta}(\ks)]}{\dif\log\ks}\,\,,
\end{equation}
For $\cs\neq 1$, we have \cite{Chen:2006nt,Cheung:2007st,Cheung:2007sv}
\begin{equation}
\label{eq:c_s-B}
P_{\zeta}(\ks)\propto\frac{H^2}{\eps\cs}\,\,,
\end{equation}
so that
\begin{equation}
\label{eq:c_s-C}
\text{$\frac{\dif\log[\ks^3P_{\zeta}(\ks)]}{\dif\log\ks} = \mathcal{O}(\eps,\eta,s)$, with $s\equiv\frac{\dot{c}_\mathrm{s}}{H\cs}\,\,.$}
\end{equation}
Approximate time translation invariance requires $s\ll1$, \ie that the inflaton sound speed does not evolve quickly in one Hubble time \cite{Cheung:2007st}. Therefore, 
$\Delta B_\zeta$ is subleading w.r.t. the bispectrum in global coordinates when $\cs\ll1$, and the $1/c_\mathrm{s}^2$-enhanced non-Gaussianity is locally observable.

\section{Connection to observations}
\label{sec:observations}

\noindent The result that we have found in \sect{bispectrum_transformation} and \sect{field_theory} can be used as 
initial condition for the study of the dynamics of small-scale perturbations in the CFC frame when they re-enter the horizon, 
which has been carried out in \cite{Dai:2015rda,Dai:2015jaa}. In this section we sketch how this can be done, leaving the details for future work. 

\begin{figure*}[!t]
\begin{center}
\includegraphics[width=0.75\columnwidth]{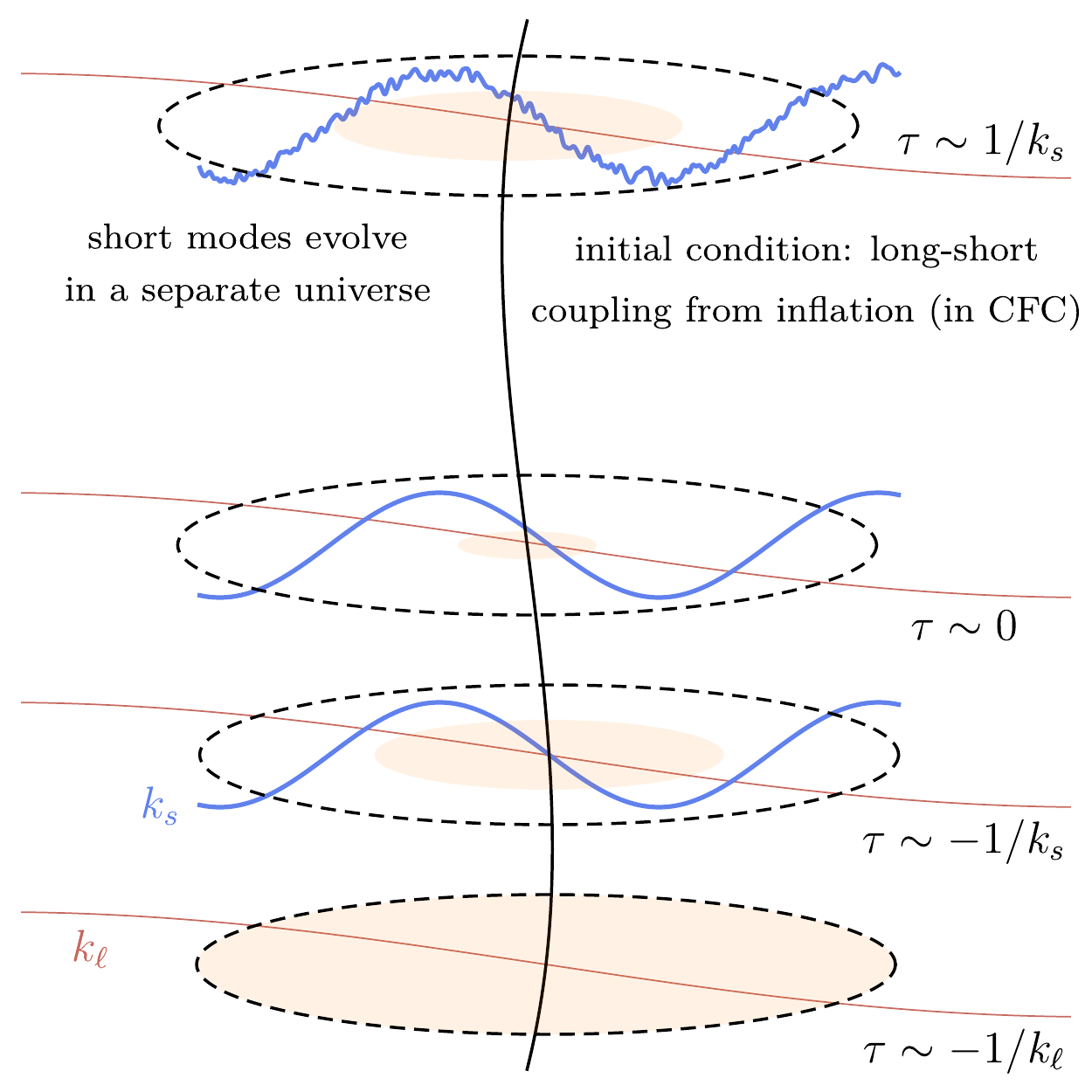} 
\end{center}
\caption{\footnotesize{For a local observer $U^\mu$, the effect of a long mode 
$\kl$ is that of making the short modes evolve in a separate universe of size $\sim 1/\kl$, 
described by a curved FLRW metric with time-dependent tidal corrections \cite{Dai:2015rda,Dai:2015jaa}. Long modes become 
classical on super-Hubble scales, and we can describe their effect on small-scale perturbations by going to CFC. We carry 
on our construction to the end of inflation, when all relevant modes are super-Hubble and time-independent. This gives the coupling between long- and short-scale perturbations 
measured by the observer $U^\mu$, that can be used as initial condition for the evolution of short modes as they re-enter the Hubble radius during the Hot Big Bang.}}
\label{fig:figure_CFC}
\end{figure*}

First, note that in order to achieve this (\ie to be able to 
use the inflationary prediction 
as initial condition for the late-time gravitational dynamics, while working in this local CFC frame throughout the whole history of short modes), the ``factoring out'' of the 
background expansion in the definition of CFC coordinates is crucial, as was already emphasized in \cite{Pajer:2013ana}. In fact, if we wanted to do the same calculation, but 
working in FNC, we could not have followed the small-scale perturbations from horizon exit to horizon re-entry: the reason is that FNC are valid on a physical 
scale $d^\mathrm{phys.}$ which is either the physical Hubble radius $H^{-1}$, or the scale of variation $a/\kl$ of long modes, whichever is smaller. During inflation all modes of interest 
exit the horizon, \ie we have $a/\kl\gg H^{-1}$ (see \fig{figure_CFC}). Hence, we have for the range of validity of FNC $d^\mathrm{phys.}\lesssim H^{-1} \ll a/k$, and FNC are therefore unable to 
cover the small-scale mode of interest with wavelength $\sim 1/\ks$.

We begin with laying out a simple procedure for how \eq{cfc_bispectrum} could in principle be measured. 
For this, we focus on the isotropic part of the long-short coupling, and assume that the long-wavelength perturbation $\zetal$ considered is 
outside the sound horizon of all fluid components. 
Then, the locally observable effects of $\zetal$ are exactly described by the separate-universe picture \cite{Dai:2015jaa}. 
Suppose now that at some time during matter domination there is a collection of comoving observers distributed throughout the Universe (\emph{e.g.}, at $z\sim 10$). 
Each observer measures the amplitude of large-scale (linear) density perturbations on a fixed physical scale $a_F/k_s$ in their local Universe, as well as their local cosmology: proper time 
since the Big Bang, Hubble rate, and spatial curvature $K = K_F$. 
Using this information and linear perturbation theory, they can immediately infer the amplitude $\mathcal{A}_s$ of the super-horizon curvature perturbations at the end of inflation at the same fixed physical scale. 
Specifically, choosing comoving gauge, they calculate the super-horizon amplitude of the perturbations $\zetas$ to their local CFC-frame metric, which, through \eq{cfc_LS_line_element}, precisely correspond to our $\zetas^F$: 
\begin{equation}
\label{eq:spatial_cfc_metric}
g_{ij}^F = \frac{a^2_F(1+2\zetas^F)}{\Big(1+\frac{K_F}{4}\abs{\vec{x}_F}^2\Big)^2}\delta_{ij}\,\,.
\end{equation}
They then communicate their local cosmology including $\mathcal{A}_s$ to a distant observer on their future light cone (\emph{e.g.}, at $z\sim 0$). 
This distant observer, now, has access to the locally measurable (in a spatial sense) amplitude of small-scale curvature perturbations $\mathcal{A}_s$ at the end of inflation, at a 
number of Lagrangian locations corresponding to Eulerian locations throughout his Hubble volume. 
He also knows what the local curvature is at each of these locations, and can use this to reconstruct the large-scale curvature perturbations $\zeta_\ell$. 
Correlating $\zeta_\ell(\vec{k}_\ell)$ with $\mathcal{A}_s(\vec{k}_\ell)$, he then obtains precisely \eq{cfc_bispectrum}, if the initial conditions are set by single-field slow-roll inflation. 

Let us now briefly discuss more realistic observables, such as the CMB temperature bispectrum \cite{Bartolo:2004if}, or the scale-dependent bias of halos \cite{Desjacques:2016bnm}. 
The contributions to any late-time observable can be split into three physically distinct contributions, as illustrated in \fig{figure_CFC}: 
\begin{itemize}[leftmargin=*]
\item \textbf{Primordial contribution:} this is defined as the contribution from \eq{cfc_bispectrum} in single-field slow-roll inflation, whose physical interpretation is given above. 
The leading contribution is $\propto k_\ell^2/k_s^2$, with a coefficient of order $\eps$, $\eta$, and of order $1/\cs^2$ for $\cs\ll 1$. Using the rough matching made 
after \eq{cfc_bispectrum}, we can approximate this as $f_\mathrm{NL}^\text{equil.} \sim \eps,\eta$, and $\sim 1/\cs^2$, respectively. 
\item \textbf{Gravitational evolution:} the gravitational dynamics that become active when the short modes re-enter the horizon contribute to the mode coupling at 
order $f_\mathrm{NL}^\text{equil.} \sim 1$ (see also \cite{Bruni:2013qta,Bruni:2014xma,Bertacca:2015mca,Bartolo:2015qva} for a discussion). Consider again an isotropic long 
mode for simplicity. Then, by way of the separate-universe picture, the leading long-short coupling can be calculated exactly by running a Boltzmann code with 
modified cosmological parameters \cite{Wagner:2014aka}. 
This contribution to the mode coupling is enhanced w.r.t. the primordial contribution for sub-horizon modes $k_s \gg \cH$, as \cite{Dai:2015jaa} has shown. During matter domination, 
the equation for the second order (\emph{i.e.} containing the long-short coupling) density contrast $\delta_{(2)}$ in CFC, for an isotropic long mode, reads (Eq.~(5.28) of \cite{Dai:2015jaa}) 
\begin{equation}
\label{eq:5_28_separate_universe}
\delta''_{(2)} + \cH\delta'_{(2)} - \frac{3}{2}\cH^2\delta_{(2)} = \frac{26}{27}\frac{1}{\cH^2}\partial^2\Phi\partial^2\phi\,\,,
\end{equation}
where $\Phi$ and $\phi$ are, respectively, the long- and short-scale Newtonian potentials, and we have used the linear (sub-Hubble) solution for $\delta_{(1)}$, that is 
\begin{equation}
\label{eq:delta_1_subhorizon}
\delta_{(1)} = \frac{2}{3\cH^2}\partial^2\phi\,\,.
\end{equation}
The initial condition from the primordial contribution, \eq{cfc_bispectrum}, for $\delta^{(2)}$, defined when the short modes re-enter the horizon ($\ks\sim\cH_\mathrm{ini}$), scales as 
\begin{equation}
\label{eq:delta_2_ini}
\delta''_{(2),\mathrm{ini}}\sim\mathcal{O}(\eps,\eta)\times\partial^2\Phi_\mathrm{ini}\frac{\partial^2\phi_\mathrm{ini}}{\cH^2_\mathrm{ini}}\,\,.
\end{equation}
The late-time evolution is hence enhanced by a factor of $\ks^2/\cH^2$, which is much larger than $1$ for sub-horizon small-scale modes. The sum 
of the two yields the late-time small-scale perturbations in the presence of the long mode in the local CFC frame. 
\item \textbf{Projection effects:} In order to connect to observations made on Earth, we have to map the CFC-frame quantities to the frame of a distant observer. 
These projection effects are calculated by following photon geodesics from the different CFC patches to the distant observer. 
Importantly, the projection effects scale as $k_\ell^2/\cH_0^2$, where $\cH_0^{-1}$ is the observer's comoving horizon. 
If $\cH_0^{-1} \gg \cH^{-1}$ like in our thought experiment above, where $\cH^{-1}$ is the comoving horizon at the time of light emission, then 
there is an interesting regime where $k_\ell \gtrsim \cH_0$ but $k_\ell \ll \cH \lesssim k_s$. 
Unlike the first two contributions above, which are suppressed by $k_\ell^2/k_s^2$ and $k_\ell^2/\cH^2$, respectively, the projection effects are not suppressed in this regime. 
They are thus the only contribution that can mimic non-Gaussianity of the local type. 
However, it is important to stress that these contributions are completely independent of the long-short coupling generated from inflation. 
They can be easily computed at linear order with the so-called ruler perturbations of \cite{Schmidt:2012ne,Jeong:2013psa,Jeong:2014ufa} (see 
also \cite{Bertacca:2014wga,Kehagias:2015tda,DiDio:2016gpd} 
for similar approaches). 
An example is provided by the squeezed-limit CMB bispectrum \cite{Creminelli:2004pv,Boubekeur:2009uk,Creminelli:2011sq,Bartolo:2011wb,Lewis:2012tc}. 
If we restrict to multipoles $\ell_\ell \lesssim \mathcal{O}(100)$, the long-wavelength mode is outside the horizon at recombination, 
so that any effect that it can have on the dynamics of short modes during recombination is suppressed, and the largest contribution comes from 
projection effects \cite{Creminelli:2004pv,Pajer:2013ana}. 
\end{itemize}

\section{Conclusions}
\label{sec:discussion}

\noindent Our main result, \eq{cfc_bispectrum}, is the three-point correlation between the large-scale curvature perturbation and the short-scale 
curvature power spectrum in Conformal Fermi Coordinates. 
This coordinate system allows us to follow the evolution of short modes in the background perturbed by the long-wavelength mode from the end of inflation until the long 
mode starts evolving again. \eq{cfc_bispectrum} encodes the primordial mode coupling that a local observer measures before it is 
reprocessed by the late-time non-linear gravitational evolution. 
We find that the magnitude of the physically relevant part of the curvature bispectrum in models of canonical single-field inflation is controlled by both $\eps$ and $\eta$ so, 
barring cancellations, the minimal amount of primordial non-Gaussianity which arises from gravitational interactions during inflation is bounded from below 
by the measured tilt of the power spectrum $n_\mathrm{s}-1$. 

As a byproduct of the calculation, we show explicitly that for $\cs < 1$ the size of non-Gaussianity is of order $(1-\cs^2)/\cs^2$ \cite{Chen:2006nt,Creminelli:2013cga}, as expected. 
The transformation to the conformal Fermi frame is proportional to $\eps$, $\eta$, and $\dot{c}_\mathrm{s}/H\cs$, and can be neglected for a slowly varying $\cs$. 
For a very small speed of sound, in fact, we do not expect gravity to play a role: the equivalence principle still demands that the bispectrum 
starts $\propto (\kl^2/\ks^2)P_{\zeta}(\ks)P_{\zeta}(\kl)$ in the squeezed limit, but the overall amplitude is fixed by inflaton derivative self-interactions.

We trace the presence of $\eta$ in our final result to the fact that it is also appearing in the cubic action $S_{(3)}$ of curvature perturbations \cite{Maldacena:2002vr}, 
\ie $\eta$ is \emph{also} a measure of the gravitational interactions of $\zeta$ during inflation. When $S_{(3)}$ is integrated by parts to show that it must have at least 
a factor of $\eps$ suppressing the interactions, a term $\propto\eps\eta$ is also introduced. 

Concerning the measurability of this effect, we see that single-field slow-roll inflation does not produce any $\fnl$ of the local type, but is guaranteed to 
produce non-Gaussianity roughly corresponding to an equilateral amplitude of $\fnl^\text{equil.} \sim 0.1\times (n_\mathrm{s}-1)$. Notice that, as discussed after \eq{cfc_bispectrum}, our 
results strictly apply to the $\mathcal{O}(k_\ell^2/k_s^2)$ part of the 
locally observable mode coupling, and hence cannot be matched unambiguously to equilateral non-Gaussianity. As discussed in \sect{observations}, this effect is 
swamped by late-time gravitational non-linearities, 
which give $\fnl^\text{equil.}$ of order $1$. It would be interesting to study models that exhibit a peculiar behavior in the squeezed limit, 
such as resonant non-Gaussianity \cite{Flauger:2010ja}, to see if they predict signatures that can be distinguished more easily from the gravitational ones. 
We leave this, 
along with the details of the connection to observations, 
to a future work.

\acknowledgments

\noindent We would like to thank Valentin Assassi, Lorenzo Bordin, Paolo Creminelli, Liang Dai, Marko Simonovi\'{c}, Gabriele Trevisan and Matias Zaldarriaga for useful 
discussions. We would also like to thank Paolo Creminelli, Alessandro Melchiorri and Marko Simonovi\'{c} for careful reading of the manuscript and useful comments. 
Tensorial algebra has partially been performed with {\tt xPand}~\cite{Pitrou:2013hga}. 
G.C. is supported by the research grant Theoretical Astroparticle Physics number 2012CPPYP7, under the program PRIN 2012 funded 
by MIUR and by TASP (iniziativa specifica INFN). 
E.P. is supported by the Delta-ITP consortium, a program of the Netherlands organization for scientific research (NWO) that is funded by 
the Dutch Ministry of Education, Culture and Science (OCW). F.S. acknowledges support from the Marie Curie Career 
Integration Grant (FP7-PEOPLE-2013-CIG) ``FundPhysicsAndLSS,'' and Starting Grant (ERC-2015-STG 678652) ``GrInflaGal'' from the European 
Research Council. G.C. would like to thank both the Delta-ITP consortium/Utrecht University and the Max Planck Institute for Astrophysics 
for the support and hospitality during two visits that have led to \mbox{this work.}

\appendix

\section{Details of the transformation to CFC}
\label{sec:details_for_cfc}

\noindent In this appendix we review in more detail the 
transformation from global coordinates to CFC, following closely the results of \cite{Dai:2015rda,Dai:2015jaa} but focusing on the 
comoving gauge for 
single-field slow-roll inflation: 
\begin{itemize}[leftmargin=*]
\item we explicitly compute the coefficients $c_n^\mu(P)$ of \eq{cfc_exponential_map}, highlighting the simplifications that occur when working at linear 
order in perturbations. We focus particularly on $c_0^\mu(P) = x^\mu(P)$, \ie the CFC coordinates of the central observer's worldline;
\item we use the results of the previous point to compute the Riemann tensor of the conformal metric on the central geodesic, and then arrive at the expression for the long-wavelength 
CFC metric. We also list the various residual gauge freedoms that are present after this step of the CFC construction;
\item we fix the freedom in the initial time used to define the CFC and the 
arbitrary constant 
that comes from the integration of the local Hubble rate; 
\item finally, we discuss the possibility of changing spatial coordinates 
without changing the time-time and time-space components of $g_{\mu\nu}^F(x_F)$: following \cite{Dai:2015jaa}, we fix 
this ambiguity by choosing a 
frame where the effect of a long-wavelength $\zeta$ on the curvature of spatial slices is explicit (we basically use the stereographic parameterization of a curved, homogeneous 
space). The freedom in the definition of the space-like vectors of the tetrad, $(e_i)^{\mu}$, \ie the choice of the integration constant in \eq{geodesic_eq_F_solved}, is discussed in 
detail in the main text (\sect{cfc_metric_inflation}). 
\end{itemize}
Before proceeding, notice that in this appendix we will use 
$\bx$ and not $x_F$ to define the CFC coordinates: this is done to simplify the notation. We will also take the CFC spatial coordinates of the central geodesic to be, generically, $\bar{\vec{x}}_c$ (instead of $\vec{0}$).

\subsection{CFC exponential map at linear order in perturbations}
\label{sec:cfc_exponential_map_at_linear_order}

\noindent Dai, Pajer and Schmidt derived the general expression for the coefficients $c_n^\mu(P)$ of \eq{cfc_exponential_map} in terms of the 
Christoffel symbols $\tilde{\Gamma}_{\mu\nu}^{\rho}$ of the conformal metric $\tilde{g}_{\mu\nu}(x) = a^{-2}_F(x) g_{\mu\nu}(x)$ evaluated on the 
central geodesic \cite{Dai:2015rda}. Up to third order in powers of $\vbx$, the transformation is
\begin{equation}
\label{eq:coordinate_transformation-1502-A}
\begin{split}
&x^\mu(\bar{\tau},\vbx) = x^\mu(P) + a_F(P)(e_i)^\mu_P\,\Delta\bar{x}^i - \frac{a^2_F(P)}{2}\tilde{\Gamma}^\mu_{\alpha\beta}|_P(e_i)^\alpha_P(e_j)^\beta_P\,\Delta\bar{x}^i\Delta\bar{x}^j \\
&\hphantom{x^\mu(\bar{\tau},\vbx) =}-\frac{a^3_F(P)}{6}(\partial_\gamma\tilde{\Gamma}^\mu_{\alpha\beta} - 2\tilde{\Gamma}^\mu_{\sigma\alpha}\tilde{\Gamma}^\sigma_{\beta\gamma})|_P(e_i)^\alpha_P(e_j)^\beta_P(e_k)^\gamma_P\,\Delta\bar{x}^i\Delta\bar{x}^j\Delta\bar{x}^k\,\,.
\end{split}
\end{equation}
In this equation, we have denoted $\bar{\vec{x}} - \bar{\vec{x}}_c$ as $\Delta\bar{\vec{x}}$, where $\bar{\vec{x}}_c$ is the CFC position of the central geodesic. Besides, we stress that all quantities are evaluated in the 
global 
coordinate system, on the central geodesic. 
For example, we have 
\begin{equation}
\label{eq:x_of_P_definition}
\text{$(e_i)^\mu_P\equiv (e_i)^\mu(x(P))$, where $x^\mu(P)= x^\mu(\bar{\tau},\vbx_c)$}\,\,.
\end{equation}
For this reason, in order to express everything in terms of barred coordinates $\bar{x}$, we need to compute $x^\mu(P)$ in terms 
of $\bar{\tau}$ (and $\bar{\vec{x}}_c$). In \cite{Dai:2015rda} it is shown that $x^\mu(\bar{\tau},\vbx_c)$ satisfies the 
equations
\begin{equation}
\label{eq:coordinate_transformation-1502-B}
\prt{x^\mu(\bar{\tau},\vbx_c)}{\bar{\tau}} = 
a_F(P)(e_0)^\mu_P\,\,,
\end{equation}
which can be easily solved if we work in perturbation theory. 
We start from 
$\mu = i$: from \eq{U_mu-flrw} we see that (keeping the notation a bit heavy for the moment)
\begin{equation}
\label{eq:coordinate_transformation-1502-B-2-a}
(e_0)^i_P = a^{-1}(\tau(\bar{\tau},\vbx_c))V^i(x(\bar{\tau},\vbx_c))\,\,,
\end{equation}
while \eq{a_F_over_a-A} reads as 
\begin{equation}
\label{eq:a_F_over_a-killer_notation}
\frac{a_F(P)}{a(P)} = 1 + C_{a_F}(\tau_\ast,\vec{x}_c(\tau_\ast)) + \int_{\tau_\ast}^\tau\dif s\,\bigg(\partial_0\zeta(s,\vec{x}_c(s)) + \frac{1}{3}\partial_i V^i(s,\vec{x}_c(s))\bigg)\,\,.
\end{equation}
As 
explained in \sect{cfc_metric_inflation}, both l.h.s. and r.h.s. of this equation are understood to be computed 
in global coordinates along the central geodesic (\ie on $\vec{x} = \vec{x}_c(\tau)$: we parameterize the central geodesic with $\tau$). Besides, we also recall that:
\begin{itemize}[leftmargin=*]
\item the first order perturbation $C_{a_F}$ is the constant coming from the integration of \eq{a_F_local_hubble};
\item $\tau_\ast$ is the initial time for the definition of CFC. 
\end{itemize} 
Before inserting this relation for the $\mu = i$ component of \eq{coordinate_transformation-1502-B}, we need to express the r.h.s. in barred 
coordinates: however, we note that $(e_0)^i_P$ is already first order in perturbations, so that the zeroth order of $a_F/a$ (which is 
$\equiv 1$) suffices. Therefore, we find 
\begin{equation}
\label{eq:coordinate_transformation-1502-B-2-b}
a_F(P)(e_0)^i_P= V^i(x(\bar{\tau},\vbx_c))\Rightarrow x^i(\bar{\tau},\vbx_c) = \bar{\vec{x}}_c + \int_{\bar{\tau}_\ast}^{\bar{\tau}}\dif \bar{s}\,V^i(\tau(\bar{s},\bar{\vec{x}}_c),\bar{\vec{x}}_c)\,\,,
\end{equation}
where we used the 
fact that $\vec{x}_c = \bar{\vec{x}}_c$ at zeroth order in perturbations. We now move to 
$\mu = 0$ in \eq{coordinate_transformation-1502-B}: using \eq{U_mu-flrw} (that defines the components of $U^\mu$ in global coordinates) we arrive at 
\begin{equation}
\label{eq:coordinate_transformation-1502-B-2-c}
a_F(P)(e_0)^0_P = \frac{a_F(P)}{a(P)}\big[1-N_1(x(P))\big] = 1 - N_1(x(P)) + \frac{a_F(P)}{a(P)}\bigg|_\ell\,\,,
\end{equation}
where we 
called $a_F(P)/a(P)|_\ell$ the first order term in \eq{a_F_over_a-killer_notation}. Inserting this into \eq{coordinate_transformation-1502-B} and 
integrating in $\bar{\tau}$ (choosing $\bar{\tau}_\ast = \tau_\ast$), we see that along the central geodesic $\tau$ is equal to $\bar{\tau} + \Delta\tau$, where $\Delta\tau$ 
is first order in perturbations. Therefore, we can 
simplify \eq{coordinate_transformation-1502-B-2-c} into
\begin{equation}
\label{eq:coordinate_transformation-1502-B-2-d}
x^i(\bar{\tau},\vbx_c) = \bar{\vec{x}}_c + \int_{\bar{\tau}_\ast}^{\bar{\tau}}\dif \bar{s}\,V^i(\bar{s},\bar{\vec{x}}_c)\,\,,
\end{equation}
and we can write \eq{a_F_over_a-killer_notation} in CFC coordinates as
\begin{equation}
\label{eq:a_F_over_a-killer_notation-cfc}
\frac{a_F(P)}{a(P)} = 1 + C_{a_F}(\bar{\tau}_\ast,\bar{\vec{x}}_c) + \int_{\bar{\tau}_\ast}^{\bar{\tau}}\dif \bar{s}\,\bigg(\partial_0\zeta(\bar{s},\bar{\vec{x}}_c) + \frac{1}{3}\partial_i V^i(\bar{s},\bar{\vec{x}}_c)\bigg)\,\,.
\end{equation}
Finally, we write the time shift $\Delta\tau$ as
\begin{equation}
\label{eq:Delta_tau}
\Delta\tau(\bar{\tau},\vbx_c) = \int_{\bar{\tau}_\ast}^{\bar{\tau}}\dif \bar{s}\big[(a_F/a)(\bar{s},\bar{\vec{x}}_c)|_\ell - N_1(\bar{s},\bar{\vec{x}}_c)\big]\,\,.
\end{equation}

Having found the expression of $x^\mu(P)$, we can move to the additional terms in \eq{coordinate_transformation-1502-A}, \ie the ones away from the central geodesic. We see that they all involve the connection coefficients (in global coordinates) of the conformal metric $\tilde{\Gamma}$, evaluated on the central geodesic. An important simplification, then, arises: since $a_F$ is equal to $a$ at zeroth order in perturbations, the conformal metric $\tilde{g}_{\mu\nu}(x) = a^{-2}_F(x)g_{\mu\nu}(x)$ will be equal to $\eta_{\mu\nu}$ at zeroth order in perturbations. 
Then, the 
Christoffel symbols will be already first order in perturbations, and \eq{coordinate_transformation-1502-A} simplifies into
\begin{equation}
\label{eq:coordinate_transformation-1502-C}
\begin{split}
&x^\mu(\bar{\tau},\vbx) = x^\mu(P) + a_F(P)(e_i)^\mu_P\,\Delta\bar{x}^i - \frac{1}{2}\tilde{\Gamma}^\mu_{ij}|_P\,\Delta\bar{x}^i\Delta\bar{x}^j - \frac{1}{6}(\partial_k\tilde{\Gamma}^\mu_{ij})|_P\,\Delta\bar{x}^i\Delta\bar{x}^j\Delta\bar{x}^k\,\,,
\end{split}
\end{equation}
where we used $a_F = a$ and $(e_i)^\mu = a^{-1}\delta^\mu_i$ at zeroth order. 
The quickest way to compute the connection coefficients of $\tilde{g}_{\mu\nu}$ is to use the relation 
\begin{equation}
\label{eq:conformal_gammas}
\tilde{\Gamma}^\rho_{\mu\nu} = \Gamma^\rho_{\mu\nu} + \delta^\rho_\mu\nabla_\nu\log\omega + \delta^\rho_\nu\nabla_\mu\log\omega - g_{\mu\nu}g^{\rho\sigma}\nabla_\sigma\log\omega\,\,,
\end{equation}
for $\tilde{g}_{\mu\nu} = \omega^2g_{\mu\nu}$.\footnote{See, \emph{e.g.}, \cite{carroll:book}.} For $\omega = a_F^{-1}$, we can 
use of the results of \cite{Dai:2015rda} for the derivatives of $a_F$ along the central geodesic, \ie 
\begin{equation}
\label{eq:covariant_derivative_of_a_F}
\text{$(\nabla_\mu \log a_F)|_P = -\frac{\cH_F(P)}{a_F(P)}(e_0)_{\mu,P}$, with $(e_0)_{\mu,P} = (g_{\mu\nu}(e_0)^\nu)_P$}\,\,.
\end{equation}
In this expression, the local comoving expansion rate $\cH_F$ is given by (again, we refer to \cite{Dai:2015rda} for details) 
\begin{equation}
\label{eq:cH_F_over_a_F}
\frac{\cH_F(P)}{a_F(P)} = \frac{1}{a(\tau)}\bigg(\cH(\tau) - \cH(\tau)N_1(s,\vec{x}_c(s)) + \partial_0\zeta(s,\vec{x}_c(s)) + \frac{1}{3}\partial_i V^i(s,\vec{x}_c(s))\bigg)\,\,,
\end{equation}
where, as in \eq{a_F_over_a-killer_notation} above, both sides of the equation are computed in global coordinates along the central geodesic. From these equations, we see that $(\nabla_\mu \log a_F)|_P$ contains also terms that are of zeroth order in perturbations. However, these terms will identically cancel with the zeroth order ones of $\Gamma^\rho_{\mu\nu}$: therefore we can safely drop the first order time shift 
and the first order shift of the position of the central geodesic 
in the argument of the Christoffel symbols 
$\tilde{\Gamma}^\rho_{\mu\nu}$. We have collected these coefficients in \tab{conformal_christoffels}: 
we note that no time derivative of the Christoffel symbols appear in \eq{coordinate_transformation-1502-C}, so we can just take their spatial derivatives in global coordinates and compute them at $(\bar{\tau},\vbx_c)$. In the end, the full transformation at order $(\bar{x}^i)^3$ reads
\begin{subequations}
\label{eq:full_transformation-A}
\begin{align}
&\tau(\bar{\tau},\vbx) = \bar{\tau} + \Delta\tau(\bar{\tau},\vbx_c) + F_i(\bar{\tau},\vbx_c)\,\Delta\bar{x}^i \nonumber \\
&\hphantom{\tau(\bar{\tau},\vbx) =} - \frac{1}{2}\tilde{\Gamma}^0_{ij}(\bar{\tau},\vbx_c)\,\Delta\bar{x}^i\Delta\bar{x}^j - \frac{1}{6}\partial_k\tilde{\Gamma}^0_{ij}(\bar{\tau},\vbx_c)\,\Delta\bar{x}^i\Delta\bar{x}^j\Delta\bar{x}^k\,\,, \label{eq:full_transformation-A-1} \\
&x^l(\bar{\tau},\vbx) = \bar{x}^l + \int_{\bar{\tau}_\ast}^{\bar{\tau}}\dif \bar{s}\,V^l(\bar{s},\bar{\vec{x}}_c) + \big[(a_F/a)(\bar{\tau},\vbx_c)|_\ell - \zetal(\bar{\tau},\vbx_c)\big]\Delta\bar{x}^l \nonumber \\
&\hphantom{x^l(\bar{\tau},\vbx) =} - \frac{1}{2}\tilde{\Gamma}^l_{ij}(\bar{\tau},\vbx_c)\,\Delta\bar{x}^i\Delta\bar{x}^j - \frac{1}{6}\partial_k\tilde{\Gamma}^l_{ij}(\bar{\tau},\vbx_c)\,\Delta\bar{x}^i\Delta\bar{x}^j\Delta\bar{x}^k\,\,. \label{eq:full_transformation-A-2}
\end{align}
\end{subequations}
where we have used the fact that 
$e^l_i$ is equal to $a^{-1}(1 - \zetal)\delta^l_i$ 
to cancel the $\bar{\vec{x}}_c$ coming from \eq{coordinate_transformation-1502-B-2-d}. In \eq{full_transformation-A-1}, we denote the sum $V_i + N_i$ as $F_i$. This definition is particularly convenient: in fact the parallel transport equation for $V^i$ in global coordinates reads 
as 
\begin{equation}
\label{eq:geodesic_eq_V-appendix}
\partial_0V^i +\mathcal{H}V^i = -\partial^i N_1 - \partial_0N^i - \mathcal{H}N^i\,\,.
\end{equation}
So, if we take $V^i$ to be $-N^i + \partial_i\digamma$, \eq{geodesic_eq_V-appendix} is solved if $\partial_0\digamma + \mathcal{H}\digamma = -N_1$, 
\ie
\begin{equation}
\label{eq:geodesic_eq_F_solved-appendix}
\digamma(x) = e^{-\int_{\tau_\ast}^\tau\dif s\,\mathcal{H}(s)}\bigg[\tau_\ast C_{\digamma}(\tau_\ast,\vec{x}) - \int_{\tau_\ast}^\tau\dif s\,e^{\int_{\tau_\ast}^s\dif w\,\mathcal{H}(w)}N_1(s,\vec{x})\bigg]\,\,,
\end{equation}
where the integration constant $C_{\digamma}$ 
is first order in perturbations. 

Now, 
to avoid having to carry 
around signs and factorials, and to simplify a little bit the notation, we rewrite Eqs.~\eqref{eq:full_transformation-A} as
\begin{equation}
\label{eq:full_transformation-B}
\begin{split}
&x^\mu(\bar{\tau},\vbx) = \bar{x}^\mu + \xi^\mu(\bar{\tau},\vbx) \\
&\hphantom{x^\mu(\bar{\tau},\vbx)}= \bar{x}^\mu + \xi^\mu(\bar{\tau},\vbx_c) + A^\mu_i(\bar{\tau},\vbx_c)\,\Delta\bar{x}^i \\
&\hphantom{x^\mu(\bar{\tau},\vbx) =}+ B^\mu_{ij}(\bar{\tau},\vbx_c)\,\Delta\bar{x}^i\Delta\bar{x}^j + C^\mu_{kij}(\bar{\tau},\vbx_c)\,\Delta\bar{x}^i\Delta\bar{x}^j\Delta\bar{x}^k\,\,.
\end{split}
\end{equation}

\begin{table}[!hbt]
\begin{center}
\begin{tabular}{lcc}
\toprule
\horsp
\vertsp $\Gamma(\eta_{\mu\nu}+h_{\mu\nu})$ \vertsp $C(a^{-1}_F) + C(a)$ \\
\hline
\moremorehorsp
${}^0_{00}$ \vertsp $\partial_0 N_1$ \vertsp $-\partial_0\zeta - \partial_m V^m/3$ \\
\morehorsp
${}^0_{0i}$ \vertsp $\partial_i N_1$ \vertsp $\mathcal{H}F_i$ \\
\morehorsp
${}^0_{ij}$ \vertsp $\partial_0\zeta\delta_{ij} - \partial_{(i}N_{j)}$ \vertsp $-(\partial_0\zeta + \partial_m V^m/3)\delta_{ij}$ \\
\morehorsp
${}^k_{00}$ \vertsp $\partial_0 N^k + \partial^k N_1$ \vertsp $\mathcal{H} F^k$ \\
\morehorsp
${}^k_{0i}$ \vertsp $\partial_0\zeta\delta_{ik} + \partial_{[i}N_{k]}$ \vertsp $-(\partial_0\zeta + \partial_m V^m/3)\delta^k_i$ \\
\morehorsp
${}^k_{ij}$ \vertsp $-\partial^k\zeta\delta_{ij} + \partial_i\zeta\delta^k_j + \partial_j\zeta\delta^k_i$ \vertsp $\mathcal{H}(-F^k\delta_{ij} + F_i\delta^k_j + F_j\delta^k_i)$ \\
\bottomrule
\end{tabular}
\caption{\footnotesize{In this table we collect the Christoffel coefficients of the conformal metric along the central geodesic in global coordinates, that we computed making use of \eq{covariant_derivative_of_a_F}. We separate them into 
the contributions from $\eta_{\mu\nu} + h_{\mu\nu}$ and 
those from the conformal factor $a^2/a^2_F$. As explained in the main text, there is no need to consider the time shift and the shift of the position of the central geodesic 
in their argument, so we omitted them. 
$F^i = \partial^i\digamma$ is defined in \eq{geodesic_eq_F_solved-appendix}: since 
it 
is a first order perturbation, we can neglect the shift 
in its argument as well.}}
\label{tab:conformal_christoffels}
\end{center}
\end{table}

\subsection{Conformal Riemann tensor and CFC (long-wavelength) metric}
\label{sec:conformal_riemann_and_cfc_long_metric}

\noindent We are now ready to compute the long-wavelength metric in the conformal Fermi frame, for which we need the conformal Riemann tensor in CFC coordinates. Since this will be already first order in perturbations, it is sufficient to calculate it in global coordinates on the central geodesic.\footnote{That is, in the definition of \eq{riemann_global_cfc} one can take the CFC coordinate basis along the central geodesic, $(\tilde{e}_\nu)^\mu_P = a_F(P)(e_\nu)^\mu_P$, at zeroth order. Using $a_F = a$ 
one remains with $(\tilde{e}_\nu)^\mu_P = \delta^\mu_\nu$. $\tilde{R}_{\mu\rho\nu\sigma}$ will not carry any power of the background scale factor by itself.} 
The calculation goes as follows: we use the 
properties of the Riemann tensor under a conformal transformation, \ie \cite{carroll:book}
\begin{equation}
\label{eq:conformal_transformation_of_riemann_tensor}
\begin{split}
&\tilde{R}^\rho_{\sigma\mu\nu} = R^\rho_{\sigma\mu\nu} - 2(\delta^\rho_{[\mu}\delta^\alpha_{\nu]}\delta^\beta_\sigma - g_{\sigma[\mu}\delta^\alpha_{\nu]}g^{\rho\beta})\nabla_\alpha\nabla_\beta\log\omega \\ 
&\hphantom{\tilde{R}^\rho_{\sigma\mu\nu} =}+2(\delta^\rho_{[\mu}\delta^\alpha_{\nu]}\delta^\beta_\sigma - g_{\sigma[\mu}\delta^\alpha_{\nu]}g^{\rho\beta} + g_{\sigma[\mu}\delta^\rho_{\nu]}g^{\alpha\beta})\nabla_\alpha\log\omega\nabla_\beta\log\omega\,\,,
\end{split}
\end{equation}
where we will take again $\omega = a_F^{-1}$. 
It is clear that if we want to compute $\tilde{R}^\rho_{\sigma\mu\nu}$ we need to know also the second (covariant) derivatives $(\nabla_\mu\nabla_\nu\log a_F)|_P$ on the central geodesic (whose zeroth order will exactly cancel the corresponding contribution from the background scale factor $a$). The coordinate-free expression for these derivatives of the local scale factor has been derived in \cite{Dai:2015rda}, and reads as
\begin{equation}
\label{eq:double_covariant_derivative_of_a_F}
\begin{split}
&(\nabla_\mu\nabla_\nu \log a_F)|_P = - \bigg(\frac{\cH_F(P)}{a_F(P)}\bigg)^2g_{\mu\nu}|_P \\ 
&\hphantom{(\nabla_\mu\nabla_\nu \log a_F)|_P =}+ \bigg[\frac{1}{a^2_F(P)}\D{\cH_F(P)}{\bar{\tau}} - 2\bigg(\frac{\cH_F(P)}{a_F(P)}\bigg)^2\bigg](e_0)_{\mu,P}(e_0)_{\nu,P}\,\,,
\end{split}
\end{equation}
where the ``local cosmic acceleration'' is given by (like in Eqs.~\eqref{eq:a_F_over_a-killer_notation}, \eqref{eq:cH_F_over_a_F}, both sides 
are understood 
as computed in global coordinates along the central geodesic)
\begin{equation}
\label{eq:cosmic_acceleration}
\frac{1}{a^2_F(P)}\D{\cH_F(P)}{\bar{\tau}} = \bigg(\frac{\cH_F(P)}{a_F(P)}\bigg)^2 + (e_0)^\mu_P\partial_\mu\bigg(\frac{\cH_F(P)}{a^2_F(P)}\bigg)\,\,.
\end{equation}

Now, as explained in \sect{cfc_metric_inflation}, we split the curvature perturbation $\zeta$ into a long- and short-wavelength part: $\zeta(x) =\zetas(x) + \zetal(x)$. Then, at leading order in $\zeta$, the metric $g_{\mu\nu}$ in global coordinates becomes
\begin{equation}
\label{eq:long_short_metric_global_coordinates-appendix}
\left.
\begin{aligned}
&\text{$g_{00} = a^2(-1-2(N_1)_s - 2(N_1)_\ell)$} \\
&\text{$g_{0i} = a^2\partial_i\psi_s + a^2\partial_i\psi_\ell$} \\
&g_{ij} = a^2(1+2\zetas + 2\zetal)\delta_{ij}
\end{aligned}
\right\}
\Rightarrow
g_{\mu\nu} = (g_{\mu\nu})_s + (g_{\mu\nu})_\ell\,\,.
\end{equation}
The goal is to absorb the effect of $\zetal$ 
by changing coordinates to CFC: therefore, we will construct the CFC metric w.r.t. $(g_{\mu\nu})_\ell$. All Christoffel symbols of \tab{conformal_christoffels}, the derivatives of the local scale factor of \eq{double_covariant_derivative_of_a_F}, and 
the 
conformal Riemann tensor can be computed in terms of $\zetal$: putting all together, and using Eqs.~\eqref{eq:cfc_metric_general-components_riemann}, we arrive at the expression for the long-wavelength metric perturbations in CFC coordinates\footnote{As discussed above, we suppress the argument $(\bar{\tau},\vbx_c)$ on the r.h.s. of these equations.}
\begin{subequations}
\label{eq:cfc_metric_general-components_riemann-explicit}
\begin{align}
&\bar{h}_{00}(\bar{\tau},\vbx) = -\Delta\bar{x}^k \Delta\bar{x}^l\bigg(\partial_k\partial_l - \frac{\delta_{kl}}{3}\partial^2\bigg)(N_1 + \partial_0\psi + \mathcal{H}\psi)\,\,, \label{eq:cfc_metric_general-components_riemann-explicit-1} \\
&\bar{h}_{0i}(\bar{\tau},\vbx) = \frac{2}{3}\Delta\bar{x}^k\Delta\bar{x}^l\bigg[(\underbrace{\partial_0\mathcal{H}-\mathcal{H}^2}_{\hphantom{\eps\cH^2\,} = \,\eps\cH^2})\big[\delta_{kl}F_i - \delta_{ki}F_l\big] 
- (\delta_{kl}\partial_i - \delta_{ki}\partial_l)(\underbrace{\mathcal{H} N_1 - \partial_0\zeta}_{\hphantom{0\,}=\,0})\bigg]
\,\,, \label{eq:cfc_metric_general-components_riemann-explicit-2} \\
&\bar{h}_{ij}(\bar{\tau},\vbx) = -\frac{1}{3}\Delta\bar{x}^k\Delta\bar{x}^l\bigg[\frac{2}{3}\mathcal{H}(\partial_m V^m)T_{ijkl} + S_{ijkl}(\zeta + \mathcal{H}\psi)\bigg]
\,\,, \label{eq:cfc_metric_general-components_riemann-explicit-3}
\end{align}
\end{subequations}
where
\begin{subequations}
\label{eq:tensors_for_cfc_metric}
\begin{align}
&T_{ijkl} = \delta_{il}\delta_{kj} - \delta_{ij}\delta_{kl}\,\,, \label{eq:tensors_for_cfc_metric-1} \\
&S_{ijkl} = \delta_{il}\partial_j\partial_k - \delta_{kl}\partial_i\partial_j + \delta_{kj}\partial_i\partial_l - \delta_{ij}\partial_l\partial_k\,\,. \label{eq:tensors_for_cfc_metric-2}
\end{align}
\end{subequations}
The last ingredient is the local scale factor $a_F(P)$: it is given by \eq{a_F_over_a-killer_notation-cfc}, \ie
\begin{equation}
\label{eq:a_F_full-appendix}
\begin{split}
a_F(\bar{\tau}) &= a(\bar{\tau} + \Delta\tau(\bar{\tau}))\bigg(1 + C_{a_F}(\bar{\tau}_\ast) + \int_{\bar{\tau}_\ast}^{\bar{\tau}}\dif \bar{s}\,\bigg(\partial_0\zeta(\bar{s}) + \frac{1}{3}\partial_i V^i(\bar{s})\bigg)\,\,,
\end{split}
\end{equation}
where we suppressed the label $\bar{\vec{x}}_c$ for simplicity.

\subsection{Fixing the residual freedom in the construction}
\label{sec:residual_gauge_freedoms_and_gauge_fixing}

\noindent In this section we discuss the additional ``gauge'' degrees of freedom present in the construction of the CFC metric. We start from the choice of initial time $\bar{\tau}_\ast$, and the constant $C_{a_F}$ in the definition of $a_F$. We will be interested in computing equal-time correlation functions as $\bar{\tau}\to 0^{-}$ (that is, on super-Hubble scales: in this way, the long modes will have have already exited the horizon, and will be classical variables that we can use in a coordinate transformation). Now, as discussed in \sect{cfc_metric_inflation}, we choose also the initial time to be $\bar{\tau}_\ast\to 0^-$. If we decide to 
fix the constant following \cite{Dai:2015rda,Dai:2015jaa}, that is by requiring that at $\bar{\tau}_\ast$ the local scale factor-proper time relation is the same as that 
of the unperturbed background cosmology, \ie
\begin{equation}
\label{eq:fix_constant_a_F-A}
\lim_{\bar{\tau}\to\bar{\tau}_\ast} a_F(\bar{\tau}) = a(\bar{\tau}_\ast)\,\,,
\end{equation}
then we see that $C_{a_F}$ 
can be safely taken 
equal to zero. In fact, expanding \eq{a_F_full-appendix} at first order in perturbations, we see that (dropping the label $\bar{\vec{x}}_c$)
\begin{equation}
\label{eq:fix_constant_a_F-B}
a_F(\bar{\tau}) = a(\bar{\tau})\big[1 + (a_F/a)(\bar{\tau})|_\ell + \cH\Delta\tau(\bar{\tau})\big]\,\,.
\end{equation}
For $\bar{\tau}$ going to zero we have that:
\begin{itemize}[leftmargin=*]
\item the integral in the definition of $\Delta\tau=\xi^0(\bar{\tau},\vbx_c)$ is killed (basically one has the limit of $x^{-1}\int_0^x\dif y\, f(y)$ for $x\to 0$), and $\cH\Delta\tau(\bar{\tau})$ 
becomes $-C_{a_F}(\bar{\tau}_\ast)$;
\item $(a_F/a)(\bar{\tau})|_\ell$, instead, simply becomes $C_{a_F}(\bar{\tau}_\ast)$.
\end{itemize}
This tells us that
for this choice of initial time $a_F$ goes to $a$ for any choice of $C_{a_F}$. Therefore we fix 
this constant to be zero in the following, for simplicity. This choice is such that $(a_F/a)(P)$, that is the difference between $a_F$ and $a$ along the central geodesic (\ie with both $a_F$ and $a$ being evaluated at the same spacetime point), goes to $1$ for $\bar{\tau}\to\bar{\tau}_\ast$.\footnote{This choice makes clear that there is no contribution from primordial physics which is not suppressed by two spatial derivatives of long-wavelength perturbations. Notice that in a curved universe the normalization of the scale factor cannot be reabsorbed by a simple rescaling of spatial coordinates. However, since $K_F$ is already first order in the long-wavelength modes, at this order any rescaling of $a_F$ can be mimicked by a coordinate transformation, and then cannot have any effect on physical observables.} 

The second gauge freedom that we discuss in this section is the possibility of changing the spatial coordinates as
\begin{equation}
\label{eq:spatial_gauge_transform_cfc-appendix}
\bar{x}^l \to \bar{x}^l(\bar{y}) = \bar{y}^l + \frac{A^l_{kij}(\bar{\tau},\vbx_c)}{6}\Delta\bar{y}^i\Delta\bar{y}^j\Delta\bar{y}^k\,\,,
\end{equation}
where the first order perturbation $A^l_{kij}(\bar{\tau},\vbx_c)$ is fully symmetric w.r.t. its three lower indices. 
Going back to $\bar{x}$ as the label for the coordinates, we see how this additional gauge freedom simply means that we can take $C^l_{kij}$ in \eq{full_transformation-B} to be not only $-\partial_k\tilde{\Gamma}_{ij}^l(\bar{\tau},\vbx_c)/6$, but 
\begin{equation}
\label{eq:full_Clkij_tensor}
C^l_{kij}(\bar{\tau},\vbx_c) = -\frac{1}{6}\big[\partial_k\tilde{\Gamma}_{ij}^l(\bar{\tau},\vbx_c) - A^l_{kij}(\bar{\tau},\vbx_c)\big]\,\,.
\end{equation}
One can show that, under 
this transformation, the CFC metric perturbations $\bar{h}_{ij}$ 
\mbox{transform as} 
\begin{equation}
\label{eq:hij_transformation-appendix}
\bar{h}_{ij}(\bar{\tau},\vbx)\to \bar{h}_{ij}(\bar{\tau},\vbx) + 
A_{(ij)kl}(\bar{\tau},\vbx_c)\Delta\bar{x}^k\Delta\bar{x}^l\,\,,
\end{equation}
where we have lowered spatial indices with $\delta_{ij}$. One can use this additional freedom to put the spatial part of the metric in the desired shape, without altering $h_{00}$ and $h_{0i}$.\footnote{Notice that, since $\bar{h}_{\mu\nu}$ is already first order in perturbations, there is no need to 
consider the change of its argument. $a_F$ will not be touched either, since it depends only on $\bar{\tau}$ which is not changed.} More precisely, we use this freedom to put the metric of Eqs.~\eqref{eq:cfc_metric_general-components_riemann-explicit} in conformal Newtonian form, following \cite{Dai:2015jaa}: we add two tensors $A^l_{kij}(\bar{\tau},\vbx)$, given by 
\begin{subequations}
\label{eq:two_tensors_Alkij}
\begin{align}
&{_{(1)}}A^l_{kij}
= -\frac{1}{6}K_F
(\delta^l_k\delta_{ij} + \delta^l_i\delta_{jk}+\delta^l_j\delta_{ki})\,\,, \label{eq:two_tensors_Alkij-1} \\
&{_{(2)}}A^l_{kij}
= \frac{1}{9}(\delta^l_k\delta_{ij} + \delta^l_i\delta_{jk}+\delta^l_j\delta_{ki})\partial^2(\zeta + \cH\psi)
\nonumber \\
&\hphantom{{_{(2)}}A^l_{kij}
=} - \frac{2}{3}(\delta^l_k\partial_i\partial_j + \delta^l_i\partial_j\partial_k + \delta^l_j\partial_k\partial_i)(\zeta + \cH\psi) \nonumber \\
&\hphantom{{_{(2)}}A^l_{kij}
=} + \frac{1}{3}(\delta_{ij}\partial^l\partial_k + \delta_{jk}\partial^l\partial_i + \delta_{ki}\partial^l\partial_j)(\zeta + \cH\psi)
\,\,, \label{eq:two_tensors_Alkij-2}
\end{align}
\end{subequations}
where we defined $K_F(\bar{\tau},\vbx)$ as 
\begin{equation}
\label{eq:curvature_K_F-appendix}
\begin{split}
K_F &= -\frac{2}{3}\big[\partial^2(\zeta + \cH\psi) + \cH\partial_m V^m\big] 
= -\frac{2}{3}(\partial^2\zeta + \cH\partial^2\digamma)\,\,.
\end{split}
\end{equation}
After this transformation, the spatial part of the metric becomes (where both l.h.s. and r.h.s. are intended as functions of $\bar{x}$)
\begin{equation}
\label{eq:spatial_part_after_stereographic_projection}
\text{$\bar{g}_{ij} = a^2_F\Bigg(\frac{1 + \Delta\bar{x}^k\Delta\bar{x}^l\mathcal{D}_{kl}(\zeta + \mathcal{H}\psi)}{\Big(1 + \frac{K_F\abs{\Delta\bar{\vec{x}}}^2}{4}\Big)^2}\Bigg)\delta_{ij}$, with $\mathcal{D}_{kl} = \partial_k\partial_l - \frac{\delta_{kl}}{3}\partial^2\,\,.$}
\end{equation}

\section{Transformation of the curvature perturbation}
\label{sec:appendix_active}

\noindent In this section we provide the transformation rules for the long- and short-wavelength curvature perturbations $\zeta$. As we have seen in \sect{zeta_transformation}, when the 
change of coordinates does not 
touch time, we can derive its effect easily with a passive approach. However, when also the time coordinate changes it is more straightforward to use an active approach. One starts from the definition of $\zeta$ given a slicing of spacetime by surfaces $\Sigma_\tau$, \ie \cite{Maldacena:2002vr,Rigopoulos:2003ak,Lyth:2004gb,Weinberg:2008nf,Weinberg:2008si,Assassi:2012et,Dias:2014msa}
\begin{equation}
\label{eq:zeta_delta_N_definition-appendix}
\zeta = \frac{\log\det(g_{ij}/a^2)}{6}\,\,,
\end{equation}
where $g_{ij}$ is the induced metric on $\Sigma_\tau$. This is nothing else but the ``$\delta N(x)$'' (local number of e-folds) definition, that relates $\zeta$ to the volume element on the $\Sigma_\tau$ surfaces. We can then use this definition to see how $\zeta$ transforms under a long-wavelength transformation 
$x^\mu\to\bar{x}^\mu = x^\mu - \xi^\mu$, $\xi^\mu = \xi^\mu_\ell$: as usual, we will stay linear in $\xi^\mu$, but we will go up to second order in perturbations (since in the end we will want to find the induced coupling between 
long and short modes). Denoting with a bar the transformed metric, at leading order in $\xi$ we have \cite{weinberg:book}
\begin{equation}
\label{eq:Lie-A}
g_{\mu\nu} \to \bar{g}_{\mu\nu} = g_{\mu\nu} + 2\nabla_{(\mu}\xi_{\nu)} = g_{\mu\nu} + g_{\nu\rho}\nabla_{\mu}\xi^\rho + g_{\mu\rho}\nabla_{\nu}\xi^\rho + \mathcal{O}(\xi^2)\,\,,
\end{equation}
so that
\begin{equation}
\label{eq:Lie-B}
\begin{split}
\bar{g}_{ij}/a^2 = \delta_{ij} + (\underbrace{e^{2\zeta} -1}_{\hphantom{\Delta g\,}\equiv\,\Delta g})\delta_{ij} + 2\nabla_{(i}\xi_{j)}/a^2 + \mathcal{O}(\xi^2)\,\,.
\end{split}
\end{equation}
Using the relation $\log\det = \Tr\log$, and working at quadratic order in perturbations (linear in $\xi$), we obtain (the ellipsis indicates that we have dropped terms of higher order in perturbations) 
\begin{equation}
\label{eq:Lie-C}
\begin{split}
\log(\bar{g}_{ij}/a^2) &= \Delta g\delta_{ij} + 2\nabla_{(i}\xi_{j)}/a^2 - \frac{1}{2}\Delta g^2\delta_{ij} - 2\Delta g\nabla_{(i}\xi_{j)}/a^2 + \mathcal{O}(\xi^2) \\
&= 2\zeta\delta_{ij} + 2\nabla_{(i}\xi_{j)}/a^2 - 2\zeta(\partial_i\xi_j + \partial_j\xi_i + 2\cH\xi^0\delta_{ij}) + \dots\,\,. 
\end{split}
\end{equation}
What we need now is the expression for 
$\nabla_{(i}\xi_{j)}/a^2$. First of all we have that
\begin{equation}
\label{eq:Lie-D}
\begin{split}
\nabla_{i}\xi_{j}/a^2 &= g_{j\rho}\nabla_{i}\xi^\rho/a^2 = g_{j\rho}\partial_{i}\xi^\rho/a^2 + g_{j\rho}\Gamma_{i\sigma}^\rho\xi^\sigma/a^2 \\
&=\partial_{i}\xi_{j} + 2\zeta\partial_{i}\xi_{j} + N_j\partial_{i}\xi^0 + g_{j\rho}\Gamma_{i\sigma}^\rho\xi^\sigma/a^2\,\,,
\end{split}
\end{equation}
where, staying linear in $\xi$ and quadratic in perturbations, $g_{j\rho}\Gamma_{i\sigma}^\rho\xi^\sigma/a^2$ is given by (see \tab{normal_christoffels})
\begin{subequations}
\label{eq:Lie-E}
\begin{align}
&g_{jk}\Gamma_{il}^k\xi^l/a^2 = \delta_{jk}\Gamma_{il}^k\xi^l 
= -\mathcal{H}N_j\xi_i + \delta_{ij}\xi^l\partial_l\zeta - 2\xi_{[i}\partial_{j]}\zeta\,\,, \label{eq:Lie-E-1} \\ 
&g_{jk}\Gamma_{i0}^k\xi^0/a^2 = e^{2\zeta}\delta_{jk}\Gamma_{i0}^k\xi^0 = \mathcal{H}\delta_{ij}\xi^0 + 2\mathcal{H}\zeta\xi^0\delta_{ij} + \xi^0\partial_0\zeta\delta_{ij} - \partial_{[i}N_{j]}\xi^0\,\,, \label{eq:Lie-E-2} \\
&g_{j0}\Gamma_{i\sigma}^0\xi^\sigma/a^2 = N_j\Gamma_{i\sigma}^0\xi^\sigma = N_j\Gamma_{ik}^0\xi^k = N_j\mathcal{H}\delta_{ik}\xi^k = \cH \xi_iN_j\,\,. \label{eq:Lie-E-3}
\end{align}
\end{subequations}
With this, \eq{Lie-C} becomes
\begin{equation}
\label{eq:Lie-H}
\begin{split}
&\log(\bar{g}_{ij}/a^2) = 2\zeta\delta_{ij} + 2\nabla_{(i}\xi_{j)}/a^2 - 2\zeta(\partial_i\xi_j + \partial_j\xi_i + 2\cH\xi^0\delta_{ij}) + \dots \\
&\hphantom{\log(g'_{ij}/a^2)} = 2\zeta\delta_{ij} + 2\partial_{(i}\xi_{j)} + 2\mathcal{H}\xi^0\delta_{ij} + 2N_{(i}\partial_{j)}\xi^0 +2\xi^\mu\partial_\mu\zeta\delta_{ij} + \dots\,\,.
\end{split}
\end{equation}
Taking the trace, we obtain 
\begin{equation}
\label{eq:Lie-I}
\bar{\zeta} = \frac{\Tr\log(\bar{g}_{ij}/a^2)}{6} = \zeta + \frac{\partial_i\xi^i}{3} + \cH\xi^0 + \frac{N^i\partial_i\xi^0}{3} + \xi^\mu\partial_\mu\zeta\,\,.
\end{equation}

\begin{table}[!hbt]
\begin{center}
\begin{tabular}{lcc}
\toprule
\horsp
 \vertsp $C(a)$ \vertsp $\Gamma$ \\
\hline
\moremorehorsp
${}^0_{00}$ \vertsp $\cH$ \vertsp $\cH + \partial_0N_1$ \\
\morehorsp
${}^0_{0i}$ \vertsp $\cH N_i$ \vertsp $\partial_i N_1 + \cH N_i$ \\
\morehorsp
${}^0_{ij}$ \vertsp $\cH\delta_{ij} + (2\zeta - N_1)\cH\delta_{ij}$ \vertsp $\cH\delta_{ij} + (2\zeta - N_1)\cH\delta_{ij} + \partial_0\zeta\delta_{ij} - \partial_{(i} N_{j)}$ \\
\morehorsp
${}^k_{00}$ \vertsp $\cH N^k$ \vertsp $\partial_0N^k + \cH N^k + \partial^kN_1$ \\
\morehorsp
${}^k_{0i}$ \vertsp $\cH\delta^k_i$ \vertsp $\cH\delta^k_i + \partial_0\zeta\delta^k_i + \frac{1}{2}(\partial_i N^k - \partial^k N_i)$ \\
\morehorsp
${}^k_{ij}$ \vertsp $-\cH\delta_{ij}N^k$ \vertsp $\partial_i\zeta\delta^k_j + \partial_j\zeta\delta^k_i-\partial^k\zeta\delta_{ij}-\cH\delta_{ij}N^k$ \\
\bottomrule
\end{tabular}
\caption{\footnotesize{In this table we collect the Christoffel coefficients of $g_{\mu\nu}$, separating the contribution 
of the conformal factor $a^2$ from the full result. We refer to \tab{conformal_christoffels} for the contribution from $\eta_{\mu\nu} + h_{\mu\nu}$.}}
\label{tab:normal_christoffels}
\end{center}
\end{table}

Now, 
recall that we are interested in a long-wavelength transformation $\xi^\mu = \xi^\mu_\ell$, and that we want to remain linear in the long mode. Then, splitting 
both $\zeta$ and $\bar{\zeta}$ in long- and short-wavelength parts, we obtain 
\begin{subequations}
\label{eq:long_short_transformation}
\begin{align}
&\bar{\zeta}_\ell = \zetal + \frac{\partial_i\xi^i_\ell}{3} + \cH\xi^0_\ell\,\,, \label{eq:long_short_transformation-1} \\
&\bar{\zeta}_s = \zetas + \frac{N^i_s\partial_i\xi^0_\ell}{3} + \xi^\mu_\ell\partial_\mu\zetas \,\,, \label{eq:long_short_transformation-2}
\end{align}
\end{subequations}
where 
$N_i = N_i(\zeta)$ is the shift constraint at linear order in $\zeta$. This shows that the short-wavelength $\zeta$ transforms as a scalar, with an additional shift if $\xi^0$ is $\vec{x}$-dependent (as it is in our case). This shift will be of no consequence for the final bispectrum transformation, in fact it is straightforward to see that both 
$N^i_s$ and $\partial_i\xi^0_\ell$ go to zero 
on super-Hubble scales (we refer to \sect{squeezed_bispectrum-A} of the main text for more details).

\section{Bispectrum in Fourier space}
\label{sec:bispectrum_fourier}

\noindent We follow closely \cite{Pajer:2013ana} to derive the transformation of the bispectrum from global coordinates to CFC. In \sect{ps_transformation} we have seen that the change to CFC gives rise to the following terms (where we have dropped the label ``$F$'' for simplicity and we have taken $\vec{x}_c\equiv(\vec{x}_1 + \vec{x}_2)/2$). As explained in the main text, only the contributions from the change in the spatial coordinates need to be considered. 
If we call $\vec{r}\equiv\vec{x}_1 - \vec{x}_2$ and $r\equiv\abs{\R}$, they are given by 
\begin{equation}
\label{eq:Delta_B-spatial-appendix}
\begin{split}
&\Delta B_\zeta = P_{\zetal A}(\abs{\vec{x}_3 - \vec{x}_c})\,r^i\partial_l\braket{\zetas\zetas}(r) \\
&\hphantom{\Delta B_\zeta =}+ \frac{1}{4}P_{\zetal C}(\abs{\vec{x}_3 - \vec{x}_c})\,r^ir^jr^k\partial_l\braket{\zetas\zetas}(r)\,\,,
\end{split}
\end{equation}
where with $P_{\zetal X}$ we denote the cross-spectrum between the long-wavelength curvature perturbation and $X$, which denotes the two tensors $A^l_i$ and $C^l_{kij}$.

We can now compute what is the contribution of these terms when we go in Fourier space $\R \leftrightarrow\vec{k}_\mathrm{S}$ and $\x_3 - \x_c \leftrightarrow \vec{k}_\mathrm{L}$. 
As shown in \cite{Pajer:2013ana}, translational invariance allows to focus separately on the long- and short-wavelength power spectra: 
\begin{itemize}[leftmargin=*]
\item a generic $P_{\zetal X_{ijk\dots}}(\abs{\vec{x}_3 - \vec{x}_c})$ will be of the form
\begin{equation}
\label{eq:generic_long_spectrum-A}
P_{\zetal X_{ijk\dots}}(\abs{\vec{x}_3 - \vec{x}_c}) = \braket{\zetal(\vec{x}_3)\partial_{ijk\dots}\zetal(\vec{x}_c)}\,\,,
\end{equation}
so that, going to Fourier space, we get
\begin{equation}
\label{eq:generic_long_spectrum-B}
\intk{\vkl}P_{\zeta}(k_\ell)\frac{\partial^{N}}{\partial x_c^i\partial x_c^j\partial x_c^k\dots} e^{i\vkl\cdot(\vec{x}_3 - \vec{x}_c)}\,\,,
\end{equation}
where $N$ is the number of derivatives we are considering. We see that each of these derivatives $\partial/\partial x_c^n$ brings down $-i\kl^n$: 
collecting these terms together with $P_{\zeta}(k_\ell)$ gives
\begin{equation}
\label{eq:generic_long_spectrum-C}
P_{\zetal X_{ijk\dots}}(\abs{\vec{x}_3 - \vec{x}_c})\to\big[(-i\kl^i)(-i\kl^j)(-i\kl^k)\dots\big]P_{\zeta}(k_\ell)\,\,;
\end{equation}
\item the short-scale spectra can be dealt with in a similar way. More precisely, a generic term that one needs to compute is of the form 
\begin{equation}
\label{eq:generic_short_spectrum-derivs-A}
(r^ir^jr^k\dots)\partial_l\braket{\zetas\zetas}(r) = \intk{\vec{k}_s}P_{\zeta}(\ks)(r^ir^jr^k\dots)\partial_le^{i\vec{k}_s\cdot\vec{r}}\,\,,
\end{equation}
that can be rewritten as 
\begin{equation}
\label{eq:generic_short_spectrum-derivs-B}
\begin{split}
&\intk{\vec{k}_s}P_{\zeta}(\ks)(r^ir^jr^k\dots)\partial_le^{i\vec{k}_s\cdot\vec{r}} = \\
&i(-i)^N\intk{\vec{k}_s}\big[\ks^lP_{\zeta}(\ks)\big]\bigg(\frac{\partial^{N}}{\partial\ks^i\partial\ks^j\partial\ks^k\dots} e^{i\vec{k}_s\cdot\vec{r}}\bigg) = \\
&i(-i)^N(-1)^N\intk{\vec{k}_s}\bigg(\frac{\partial^{N}}{\partial\ks^i\partial\ks^j\partial\ks^k\dots}\big[\ks^lP_{\zeta}(\ks)\big]\bigg) e^{i\vec{k}_s\cdot\vec{r}}\,\,,
\end{split}
\end{equation}
where $N$ is the number of powers of $\vec{r}$ that we are considering. We have moved the derivatives from the exponential to the power spectrum integrating by parts $N$ times. This generates an overall $(-1)^N$ factor. Then, we see that the Fourier transform of $(r^ir^jr^k\dots)\partial_l\braket{\zetas\zetas}(r)$ is given by 
\begin{equation}
\label{eq:generic_short_spectrum-derivs-C}
(r^ir^jr^k\dots)\partial_l\braket{\zetas\zetas}(r)\to i^{N+1}\frac{\partial^{N}}{\partial\ks^i\partial\ks^j\partial\ks^k\dots}\big[\ks^lP_{\zeta}(\ks)\big]\,\,.
\end{equation}
\end{itemize}

For our applications, we will need to take $N$ up to $3$. The expressions can quickly become cumbersome, so we proceed step by step and collect the intermediate results for convenience of the reader. Since all derivatives $\partial/\partial\ks^i$ are acting on a function of $\ks$ only, some simplifications will arise:
\begin{itemize}[leftmargin=*]
\item we start from the simple $\partial/\partial\ks^i$, that we rewrite as 
\begin{equation}
\label{eq:derivatives_wrt_k-A}
\prt{}{\ks^i} = \frac{\ks^i}{\ks^2}\ds\,\,.
\end{equation}
This directly leads to
\begin{equation}
\label{eq:derivatives_wrt_k-B}
\prt{}{\ks^i}\ks^j = \delta^j_i + \ks^j\prt{}{\ks^i} = \delta^j_i + \frac{\ks^i\ks^j}{\ks^2}\ds\,\,;
\end{equation}
\item then we will encounter terms like $\partial^2/\partial\ks^i\partial\ks^j$. With simple manipulations one arrives at
\begin{equation}
\label{eq:derivatives_wrt_k-C}
\frac{\partial^2}{\partial\ks^i\partial\ks^j} = \frac{\delta_{ij}}{\ks^2}\ds + \frac{\ks^i\ks^j}{\ks^4}\bigg(\dsh{2} - 2\ds\bigg)\,\,;
\end{equation}
\item finally, we will have terms with three derivatives and one power of $\vks$, \ie
\begin{equation}
\label{eq:diff_operator-short}
\mathcal{S}^l_{ijk} = \frac{\partial^3}{\partial\ks^i\partial\ks^j\partial\ks^k}\ks^l\,\,. 
\end{equation}
If we define
\begin{subequations}
\label{eq:diff_operators-help}
\begin{align}
&\mathcal{D}_1 = \ds\,\,, \label{eq:diff_operators-help-1} \\
&\mathcal{D}_2 = \dsh{2} - 2\ds\,\,, \label{eq:diff_operators-help-2} \\
&\mathcal{D}_3 = \dsh{3} - 6\dsh{2} + 8\ds\,\,, \label{eq:diff_operators-help-3}
\end{align}
\end{subequations}
we can write this term as a sum of various pieces (all symmetric in $i$, $j$, $k$)
\begin{equation}
\label{eq:diff_operator-short-full}
\begin{split}
&\mathcal{S}^l_{ijk} = \frac{\delta_{ij}\delta_{kl}}{\ks^2}\mathcal{D}_1 + \text{$2$ perms.} 
+\frac{\delta_{ij}\ks^k\ks^l}{\ks^4}\mathcal{D}_2 + \text{$2$ perms.} \\
&\hphantom{\mathcal{S}^l_{ijk} =}
+\frac{\delta_{li}\ks^j\ks^k}{\ks^4}\mathcal{D}_2 + \text{$2$ perms.} 
+\frac{\ks^i\ks^j\ks^k\ks^l}{\ks^6}\mathcal{D}_3\,\,.
\end{split}
\end{equation}
\end{itemize}
With some simple algebra, 
one can now write the expression for the action of $\mathcal{S}^l_{ijk}$ on the small-scale power spectrum at leading order in slow-roll, recalling that if we 
neglect any running of the spectral index 
we can write derivatives of 
$P_\zeta(\ks)$ as
\begin{equation}
\label{eq:derivatives_wrt_logS}
\frac{\dif^mP_{\zeta}(\ks)}{\dif\log\ks^m} = (n_\mathrm{s} - 4)^mP_{\zeta}(\ks) = (-3)^m\bigg[1+\frac{m}{3}(\ns-1)\bigg]P_\zeta(\ks)\,\,.
\end{equation}

\section{Small speed of sound: overview of the calculation}
\label{sec:small_sound_speed}

\noindent In this section we investigate briefly the simplifications that arise when one is interested in the limit of a small inflaton speed of sound ($c_\mathrm{s}\ll 1$). In passing, we collect some results that can be useful if one wants to compute the CFC bispectrum directly from the action, with the long-wavelength metric given by \eqsI{cfc_metric_general-components_riemann-explicit-slow_roll}. This approach is different from the one we have followed in \sect{field_theory}, where we obtained the $\eta$ contribution to the CFC bispectrum by mirroring Maldacena's calculation in flat gauge \cite{Maldacena:2002vr}. 

We will take the Goldstone boson of time diffeomorphisms (that we will call $\pi$) as short-wavelength variable \cite{Cheung:2007st,Cheung:2007sv}. Before 
proceeding, let us see how the metric in the conformal Fermi frame looks like when working in the $\pi$ gauge for the short modes: dropping for simplicity the ``$F$'' label not only on coordinates, but also on all the components of the metric (for simplicity of notation), we have that
\begin{equation}
\label{eq:app_D-i}
\begin{split}
&\dif s^2 = -\underbrace{a^2(1 + 2 (N_1)_\ell + 2(N_1)_s)}_{\hphantom{N^2\,}=\,N^2}\dif\tau^2 + a^2
N^i_s
(\dif\tau\dif x^i + \dif x^i\dif\tau) + \underbrace{a^2e^{2\zetal}\delta_{ij}}_{\hphantom{\gamma_{ij}\,}=\,\gamma_{ij}}\dif x^i\dif x^j\,\,,
\end{split}
\end{equation}
where:
\begin{itemize}[leftmargin=*]
\item we have taken $a_F = a$ in $g_{0i}$. The reason is that we can remain at linear order in perturbations when we deal with the time-time and time-space components of the metric;
\item we have put to zero the long-wavelength shift constraint, because we have seen in \sect{cfc_metric_inflation} that it is of order $\kl^3$. Besides, the short-scale shift constraint $N^i_s$ can be written as $\partial_i\psi$, as usual: we will omit the ``$s$'' subscript in the following for simplicity of notation. We note that this definition (\emph{i.e.} without including the factor of $a^2$) agrees with the ADM parameterization of 
$g_{0i}$ (which is $\gamma_{ij}N^j$), because we are working at linear order in the 
constraints. Therefore, in the following we will raise and lower the indices of $N^i$ with $\delta_i^j$;
\item both $(N_1)_s$ and $\psi$ will be linearly solved in terms of $\pi$ \cite{Cheung:2007st,Cheung:2007sv}. In single-field slow-roll inflation, the leading interaction (cubic) Lagrangian comes from the mixing with gravity, so it is not possible to neglect these terms (\emph{i.e.}, the decoupling limit would not capture the relevant physics);\footnote{Even if, as we have seen in \sect{field_theory}, there are a lot of simplifications that arise if we are interested only in 
contributions to the bispectrum that are $\propto\eta$. 
}
\item the long-wavelength contribution to $a_F$, which is equal to (we refer to \sect{conformal_riemann_and_cfc_long_metric} for details) 
\begin{equation}
\label{eq:app_D-ii}
a_F(\tau) = a(\tau)\big[1 + (a_F/a)(\tau)|_\ell + \cH\xi^0(\tau,\vec{0})\big] \,\,,
\end{equation}
is included in $(N_1)_\ell$ and $\zetal$ (the subscript ``$\ell$'' is dropped on $\xi^\mu$ for simplicity). That is, we add it to the 
perturbations $h_{00}$ and $h_{ij}\propto\delta_{ij}$ that make up the long-wavelength CFC metric of \eqsI{cfc_metric_general-components_riemann-explicit-slow_roll}. 
In this way it is easier to keep track of both the order in perturbations and the order in the slow-roll expansion.
\end{itemize}
In this gauge, the action is equal to 
\begin{equation}
\label{eq:app_D-iii}
\begin{split}
S = S_\mathrm{EH} + \int\dif^4x\,&N\sqrt{\gamma}\bigg[\frac{(\partial_0\phi - N^i\partial_i\phi)^2}{N^2} - \gamma^{ij}\partial_i\phi\partial_j\phi - 2 V(\phi)\bigg] 
\,\,,
\end{split}
\end{equation}
where we have that: 
\begin{itemize}[leftmargin=*]
\item the inflaton $\phi$, whose background value we write as $\bar{\phi}$, is given by (in the following, we will often denote derivatives w.r.t. $\tau$ with a ``prime'') 
\begin{equation}
\label{eq:app_D-iv}
\begin{split}
\phi
&= \bar{\phi}(\tau+\pi
) + \underbrace{\bar{\phi}'(\tau+\pi
)\xi^0(\tau+\pi
)}_{\hphantom{\vphi_\ell(\tau+\pi)\,}\equiv\,\vphi_\ell(\tau+\pi)} +\,\mathcal{O}[(\xi^0)^2] \\ 
&= \sqrt{2\eps}\,\cH\bigg[\xi^0
+ \pi
+ \partial_0\xi^0
\pi
+ \frac{\cH}{2}\bigg(1-\eps+\frac{\eta}{2}\bigg)(\pi^2
+2\xi^0\pi)
\bigg] + \dots\,\,,
\end{split}
\end{equation}
where we have dropped terms cubic in perturbations (staying linear in the long-wavelength $\xi^0$) and we have used the slow-roll relations
\begin{subequations}
\label{eq:app_D-v}
\begin{align}
&\bar{\phi}' = \sqrt{2\eps}\,\cH\,\,, \label{eq:app_D-v-1} \\
&\cH' = \cH^2(1-\eps)\,\,, \label{eq:app_D-v-2} \\
&\eps' = \cH\eps\eta\,\,. \label{eq:app_D-v-3}
\end{align}
\end{subequations}
The presence of $\bar{\phi}'\xi^0\equiv\vphi_\ell$ is due to the transformation to CFC, and the fact that at second order in $\kl$ we cannot neglect the change in the time coordinate; 
\item the potential $V(\phi)$ can likewise be expanded in perturbations, using the above result for $\phi$ and the fact that $V(\bar{\phi}) = H^2(3-\eps)$. We will not write down the expansion here, since it is very 
easy to obtain it with simple algebra. We note that useful relations between $V(\bar{\phi})$ (and its derivatives) and the Hubble slow-roll parameters are also listed in Sec.~B of \cite{Pajer:2016ieg}; 
\item the Einstein-Hilbert action $S_\mathrm{EH}$, \ie 
\begin{equation}
\label{eq:app_D-vi}
S_\mathrm{EH} = \frac{1}{2}\int\dif^4x\,N\sqrt{\gamma}\bigg[R^{(3)}(\gamma) + \frac{E^{ij}E_{ij} - E^2}{N^2}\bigg]\,\,,
\end{equation}
with 
\begin{subequations}
\label{eq:app_D-vii}
\begin{align}
&E_{ij} \equiv \frac{1}{2}\big[\partial_0\gamma_{ij} - 2\nabla_{(i}N_{j)}\big]\,\,, \label{eq:app_D-vii-1} \\
&E = \gamma^{ij}E_{ij}\,\,, \label{eq:app_D-vii-2}
\end{align}
\end{subequations}
is computed in terms of the metric of \eq{app_D-i}.
\end{itemize}
It is now straightforward to solve the constraints in terms of $\pi$: at linear order in perturbations they are given by \cite{Cheung:2007sv} 
\begin{subequations}
\label{eq:app_D-viii}
\begin{align}
&(N_1)_s = \eps\cH\pi\,\,, \label{eq:app_D-viii-1} \\
&\psi = -\eps\cH\partial^{-2}\partial_0\pi\,\,. \label{eq:app_D-viii-2}
\end{align}
\end{subequations}
From this 
one 
can find the quadratic action for the Goldstone boson $\pi$. At leading order in slow-roll it is equal to \cite{Cheung:2007st,Cheung:2007sv} 
\begin{equation}
\label{eq:app_D-ix}
S_{\pi\pi} = \int\dif^4x\,a^2\cH^2\eps\big[(\partial_0\pi)^2 - (\partial_i\pi)^2\big]\,\,.
\end{equation}

Now, what we are looking for is the coupling 
between long and short modes, so what we need is the interaction Lagrangian at cubic order in perturbations with one long leg and two short ones (we focus only on scalar degrees of freedom, \ie we discard the graviton). Adding this to the quadratic action for $\pi$, one can compute the power spectrum of $\pi$ in the background of a long-wavelength 
classical 
curvature perturbation: we will denote this two-point function $\braket{\pi\pi}|_\ell$ by $P_\pi|_\ell$ 
(we use the subscript ``$\ell$'' to indicate 
that the power spectrum of $\pi$ will depend on the whole long-wavelength part of the metric in \eq{app_D-i}, \ie on $a_F$, $K_F$, etc.). 
Once the cubic action $S_{\pi\pi}|_{\ell}$ has been found, one can use the in-in formalism \cite{Schwinger:1960qe,Jordan:1986ug,Calzetta:1986ey,Maldacena:2002vr,Weinberg:2005vy,Chen:2006nt}, which guarantees the correct choice of normalization and vacuum for the modes, to calculate $P_\pi|_\ell$. Since we are computing 
a two-point function in a perturbed background 
FLRW, and not a full three-point function, there is a simplification \cite{Creminelli:2013cga}: 
the cubic Lagrangian will depend explicitly on the spatial coordinates, since the long-wavelength metric in CFC does. However, the terms coming from the correction to the scale factor are evaluated only on the central geodesic, and do not depend on $\vec{x}$: schematically, we denote these terms by $S_{\pi\pi}|_{\ell,\vec{x}=\vec{0}}$. Therefore, it is possible (but not necessary) to deal with them by taking as free action not only the one of \eq{app_D-ix}, but $S_{\pi\pi} + S_{\pi\pi}|_{\ell,\vec{x}=\vec{0}}$. The resulting equation of motion can be solved perturbatively with Green's function methods (see \cite{Pajer:2016ieg}, for example), and the normalization of the modes (necessary to have the correct commutation relations) and the choice of vacuum (\ie the Bunch-Davies vacuum) can be carried out in the usual way.\footnote{For example, for the normalization of the modes it will be necessary to impose that the Wronskian of the mode functions of the canonically normalized variable is equal to $1$ \cite{Maldacena:2002vr,Chen:2006nt,Cheung:2007sv,Baumann:2009ds,Lim:2012lectures}.} 
At this point, $P_\pi|_\ell$ is $P_\pi + P_\pi|_{\ell,\vec{x}=\vec{0}}$. To find the final contribution $P_\pi|_{\ell,\vec{x}\neq\vec{0}}$, which comes from the $\vec{x}$-dependent terms in the cubic action, one can do a tree-level in-in calculation. Denoting by $\mathcal{L}_{\pi\pi}|_{\ell,\vec{x}\neq\vec{0}}$ the corresponding cubic Lagrangian, and using the fact that at third order in perturbations the interaction Hamiltonian density is $-\mathcal{L}_\mathrm{int.}$, one can write this power spectrum on super-Hubble scales ($\tau\to0^-$) as \cite{Maldacena:2002vr,Chen:2006nt,Cheung:2007sv,Baumann:2009ds,Lim:2012lectures,Creminelli:2013cga} 
\begin{equation}
\label{eq:app_D-x}
\begin{split}
\braket{\pi(
0,\vec{x}_1)\pi(
0,\vec{x}_2)}|_{\ell,\vec{x}\neq\vec{0}} = 
(2\pi)^3\int\frac{\dif\vec{k}_1\dif\vec{k}_2}{(2\pi)^6} &\mathcal{P}_{ij}(\tau,\vec{k}_1,\vec{k}_2) \\
&\times\bigg[\frac{\partial^2}{\partial k_1^i\partial k_1^j}\delta(\vec{k}_1+\vec{k}_2)\bigg]
e^{i\vec{k}_1\cdot\vec{x}_1+i\vec{k}_2\cdot\vec{x}_2}
\,\,.
\end{split}
\end{equation}
In the above equation, the function $\mathcal{P}_{ij}$ is defined as \cite{Creminelli:2013cga} 
\begin{equation}
\label{eq:app_D-xi}
\mathcal{P}_{ij}(\tau,\vec{k}_1,\vec{k}_2) = -4\,\mathrm{Re}\bigg[i\pi(\tau,\vec{k}_1)\pi(\tau,\vec{k}_2)\int^
0_{-\infty^+}\dif s\,\mathcal{L}^\ast_{ij}(s,\vec{k}_1,\vec{k}_2)\bigg]\,\,,
\end{equation}
where the boundary condition $-\infty^+\equiv -\infty(1-i\epsilon)$ picks out the interacting vacuum. With $\mathcal{L}^\ast_{ij}$ we denote the (complex conjugate of the) Fourier transform of $\mathcal{L}_{\pi\pi}|_{\ell,\vec{x}\neq\vec{0}}$ evaluated on the mode functions of $\pi$, that we will denote by $\pi_\mathrm{cl.}$. The $i$, $j$ indices mean that every explicit power of $\vec{x}$ that is carried by the long legs (which are all quadratic in $x^i$, \eg $\propto K_F\abs{\vec{x}}^2$) is taken care of by the derivatives of $\delta(\vec{k}_1+\vec{k}_2)$ in \eq{app_D-x}. More precisely: 
\begin{itemize}[leftmargin=*]
\item suppose that $\mathcal{L}_{\pi\pi}|_{\ell,\vec{x}\neq\vec{0}}$ contains a term of the form 
\begin{equation}
\label{eq:app_D-xii}
\mathcal{L}_{\pi\pi}|_{\ell,\vec{x}\neq\vec{0}}\supset a_1\,a^2\cH^2\eps\, K_F\abs{\vec{x}}^2\,(\partial_i\pi)^2\,\,,
\end{equation}
where $a_1$ is a numerical factor. In this case, $\mathcal{L}_{ij}$ would be equal to 
\begin{equation}
\label{eq:app_D-xiii}
\mathcal{L}
_{ij}(\tau,\vec{k}_1,\vec{k}_2)\supset -a_1\times a^2\cH^2\eps \times \underbrace{K_F\delta_{ij}}_{K_F\abs{\vec{x}}^2
}\times\underbrace{(i\vec{k}_1)\cdot(i\vec{k}_2)\,\pi_\mathrm{cl.}(\tau,k_1)\pi_\mathrm{cl.}(\tau,k_2)}_{
(\partial_i\pi)^2
}\,\,.
\end{equation}
We stress that $K_F$ is just a real, classical, $\vec{x}$-independent number (it is evaluated on the central geodesic), therefore it is on the same footing as $a^2\cH^2\eps$, \ie it is not touched by the Fourier transform (and it is already evaluated on the classical mode functions). We note that the time dependence of $K_F$ starts at $\mathcal{O}(\kl^4)$, so it can be considered a constant at the order in the gradient expansion that we are working;
\item one can also consider the case where that $\mathcal{L}_{\pi\pi}|_{\ell,\vec{x}\neq\vec{0}}$ contains an anisotropic term. From \eq{cfc_metric_general-components_riemann-explicit-slow_roll-3}, we have that 
the anisotropic part of $\zetal$ is (in terms of the long-wavelength curvature perturbation in global coordinates $\zeta_\mathrm{gl.}$)
\begin{equation}
\label{eq:app_D-xiv}
\zetal^\mathrm{anis.}(x) = \frac{1}{2}x^i x^j\mathcal{D}_{ij}\big[\eps\mathcal{H}\partial^{-2}\partial_0\zeta_\mathrm{gl.}(\tau,\vec{0})\big]\equiv x^ix^jZ_{ij}(\tau,\vec{0})\,\,,
\end{equation}
with $\mathcal{D}_{ij} = \partial_i\partial_j - \partial^2\delta_{ij}/3$. Then the cubic Lagrangian will contain a term of the form ($a_2$ is again a numerical factor) 
\begin{equation}
\label{eq:app_D-xv}
\mathcal{L}_{\pi\pi}|_{\ell,\vec{x}\neq\vec{0}}\supset a_2\,a^2\cH^2\eps\, x^ix^jZ_{ij}
\, (\partial_0\pi)^2\,\,,
\end{equation}
and $\mathcal{L}
_{ij}$ would, similarly to \eq{app_D-xiii}, be given by
\begin{equation}
\label{eq:app_D-xvi}
\mathcal{L}
_{ij}(\tau,\vec{k}_1,\vec{k}_2)\supset -a_1\times a^2\cH^2\eps \times Z_{ij}
\times
\partial_0\pi_\mathrm{cl.}(\tau,k_1)\partial_0\pi_\mathrm{cl.}(\tau,k_2)
\,\,.
\end{equation}
As before, $Z_{ij}$ is a real number: however, in this case one cannot neglect its time dependence when computing the corresponding $\mathcal{P}_{ij}$, since it starts at order $\kl^2$. 
\end{itemize}
We note that an overall $-1$ in the definition of $\mathcal{P}_{ij}$ is due to the fact that 
\begin{equation}
\label{eq:app_D-xvii}
x^ix^j = -\int\frac{\dif\vec{k}}{(2\pi)^3}\bigg[\frac{\partial^2}{\partial k^i\partial k^j}\delta(\vec{k})\bigg]e^{i\vec{k}\cdot\vec{x}}\,\,,
\end{equation}
while an overall factor of $2$ comes from the two different contractions that we need to consider when we use Wick's theorem. 
Now, integrating by parts \eq{app_D-x} to 
isolate a $(2\pi)^3\delta(\vec{k}_1+\vec{k}_2)$, it is possible to extract the expression for $P_\pi|_{\ell,\vec{x}\neq\vec{0}}$. Multiplying it with a second long mode, and taking the average, gives then 
the squeezed limit bispectrum in CFC. 

Eventually, one is interested in the short-wavelength $\zetas$ and its coupling with the long mode. In unitary gauge $\pi = 0$, the perturbation $\zetas$ is defined by
\begin{equation}
\label{eq:app_D-xviii}
\gamma_{ij} = a^2e^{2\zetal}e^{2\zetas}\delta_{ij}\,\,,
\end{equation} 
so what one needs to do is find the relation between $\zetas$ and the Goldstone boson $\pi$. We see from \eq{app_D-iv} that a time shift $\tau = \tilde{\tau}-\pi$ would take care of the inflaton perturbation, that would go back to $\phi = \bar{\phi} + \vphi_\ell$ (as it was after the transformation from global coordinates in $\zeta$ gauge to CFC) at linear order in $\pi$. This is enough for our purposes, since we are interested only in the long-short coupling and therefore we can drop all terms that are quadratic (or higher) in $\pi$. 
Correspondingly, at quadratic order in perturbations, the spatial metric would transform as (see also \sect{field_theory} for more details) 
\begin{equation}
\label{eq:app_D-xix}
\begin{split}
&\hat{\tilde{g}}_{ij} = -a^2\prt{\pi}{\tilde{x}^i}\prt{\pi}{\tilde{x}^j} - a^2\prt{\pi}{\tilde{x}^i}\partial_j\psi - a^2\prt{\pi}{\tilde{x}^j}\partial_i\psi + a^2e^{2\zetal}e^{-\cH\pi}e^{-\pi\partial_0\zetal}\delta_{ij}\,\,,
\end{split}
\end{equation}
where in the expansion of $\gamma_{ij}$ (\ie the last term on the r.h.s.) we have stopped at linear order in the short mode, for the same reason discussed above. We see that the metric, after this time shift, is not of the form of \eq{app_D-xviii}, because of the first two terms that involve spatial derivatives of $\pi$. It is possible to remove them with a second order spatial coordinate transformation: however, since the terms we have to remove are of quadratic order in the short modes, their contribution to $\zetas$ would be negligible for our purposes. From this, comparing with \eq{app_D-xviii}, we conclude that the relation between $\zetas$ and $\pi$ is given by 
\begin{equation}
\label{eq:app_D-xx}
\zetas = -\cH\pi-\pi\partial_0\zetal\,\,.
\end{equation}
After the in-in calculation of $P_\pi|_\ell$ that we have briefly discussed above has been carried out, one can use \eq{app_D-xx} to compute the power spectrum of the short-scale $\zetas$ in the background of the long modes: in addition to the coupling coming from the interactions (\ie the contribution 
coming from replacing $\pi$ with $-\zetas/\cH$ in $P_\pi|_\ell$), there will be additional terms coming from the second order (long-short) term 
$\zetas\supset-\partial_0\zetal\pi$. 

At this point, one must compute the cubic Lagrangian, and for each term derive the corresponding $\mathcal{P}_{ij}$. However, since we are working in a ``mixed $\zeta$ - $\pi$ gauge'', it is clear that there will be some complications due to the fact that the interaction Lagrangian will not be slow-roll suppressed w.r.t. the quadratic Lagrangian for the pion. For example, there will be interactions of the form 
\begin{equation}
\label{eq:app_D-xxi}
\begin{split}
\mathcal{L}_{\pi\pi}|_\ell\supset \big\{&a_1\,a^2\cH^2\eps\,\zetal\,(\partial_0\pi)^2,a_2\,a^2\cH^2\eps\,\zetal\,(\partial_i\pi)^2, \\
&a_3\,a^2\cH^4\eps\,\partial_0\xi^0\,\pi^2,a_4\,a^2\cH^2\eps\,\partial_i\xi^0\partial_i\pi\,\pi,\dots\big\}\,\,,
\end{split}
\end{equation}
coming from both the EH action (once we plug in it the constraints solved in terms of $\pi$) and the inflaton action. Since the mode functions $\pi_\mathrm{cl.}$, $(\zetal)_\mathrm{cl.}$ and $\xi^0_\mathrm{cl.}$ are $\propto1/\sqrt{\eps}$ at leading order in slow-roll, the cubic Lagrangian should be at least of order $\eps^{3/2}$ to be able to capture the leading part of the bispectrum (which we know is slow-roll suppressed, \ie it is of order $(\eps^2,\eps\eta) \times (1/\sqrt{\eps}\,)^6$) by using the de Sitter modes alone. In other words, if we were to compute the 
bispectrum using the in-in formalism discussed above, we would indeed see that at zeroth order in slow-roll (that is, at order $(1/\sqrt{\eps}\,)^6$) 
it is zero, and that the leading order result is $\mathcal{O}(\eps,\eta)$. However, we could not trust the 
slow-roll-suppressed part of the result because we would be neglecting contributions coming from corrections to the mode functions: to capture all the effects it would be necessary to use the full classical solutions in terms of Hankel functions, which complicate considerably the time integrals of \eq{app_D-xi}. 

We are now in the position to discuss briefly the case of an inflaton speed of sound $c_\mathrm{s}$ different from $1$. 
We know that for $c_\mathrm{s}\neq1$, the contribution to the bispectrum which is not slow-roll suppressed will be different from zero: namely, it will be proportional to $(1-c^2_\mathrm{s})/c_\mathrm{s}^2$ \cite{Chen:2006nt,Cheung:2007sv,Creminelli:2013cga}. Therefore, the de Sitter modes would be able to fully capture the leading order 
bispectrum (which would be much larger than its slow-roll suppressed part if $c_\mathrm{s}$ is not too close to $1$) in this case. Besides, a further simplification arises if $c_\mathrm{s}\ll 1$: in fact, in this non-relativistic limit we do not expect the short modes $\pi$ to feel the spatial curvature of the universe induced by the long mode,\footnote{The same argument can be used for the anisotropic part of the long-wavelength metric (which we also know has an additional slow-roll suppression w.r.t. the other parts).} but to be sensitive only to the effect it has on the expansion history $a_F\neq a$ \cite{Creminelli:2013cga}. Translated at the level of the interaction Lagrangian, this statement means that 
it is possible to drop all the long legs that are not (functions of) $a_F$, because only these will affect the bispectrum at order $(\kl^2/\ks^2)/c^2_\mathrm{s}$ \cite{Creminelli:2013cga}. Then, powers of $x^i$ will not appear explicitly in $\mathcal{L}_{\pi\pi}|_\ell$ and it will not be necessary to compute $\mathcal{P}_{ij}$ using the method of \eq{app_D-xi}, greatly simplifying the calculation. 

\clearpage


\end{document}